\documentclass[10pt,letterpaper]{article}
\usepackage[top=0.85in,left=2.75in,footskip=0.75in]{geometry}

\usepackage{amsmath,amssymb}
\usepackage{amsmath}
\usepackage[table,x11names]{xcolor}
\usepackage[colorinlistoftodos]{todonotes}
\usepackage{hyperref}
\hypersetup{colorlinks,allcolors=blue}
\usepackage{dcolumn}
\usepackage{bm}
\usepackage{subcaption}
\usepackage{float}
\usepackage[export]{adjustbox}
\DeclareMathOperator{\Hessian}{Hess}
\newcommand{\bigzero}{\mbox{\normalfont\Large\bfseries 0}}
\newcommand*{\Scale}[2][4]{\scalebox{#1}{\ensuremath{#2}}}%
\DeclareMathOperator{\Var}{Var}

\usepackage{changepage}

\usepackage[utf8x]{inputenc}

\usepackage{textcomp,marvosym}

\usepackage{cite}

\usepackage{nameref}

\usepackage{microtype}
\DisableLigatures[f]{encoding = *, family = * }

\usepackage[table]{xcolor}

\usepackage{array}

\newcolumntype{+}{!{\vrule width 2pt}}

\newlength\savedwidth

\raggedright
\usepackage[aboveskip=1pt,labelfont=bf,labelsep=period,justification=raggedright,singlelinecheck=off]{caption}

\makeatletter
\renewcommand{\@biblabel}[1]{\quad#1.}
\makeatother

\usepackage{lastpage,fancyhdr}
\usepackage{epstopdf}
\fancyheadoffset[L]{2.25in}
\fancyfootoffset[L]{2.25in}
\begin{document}
\vspace*{0.2in}

\begin{flushleft}
{\Large
\textbf\newline{Inferring phenomenological models of first passage processes} 
}
\newline
\\
Catalina Rivera\textsuperscript{1,*},
David Hofmann\textsuperscript{1,3},
Ilya Nemenman\textsuperscript{1,2,3}

\bigskip
\textbf{1} Department of Physics, Emory University, Atlanta, Georgia, United States of America
\\
\textbf{2} Department of Biology, Emory University, Atlanta, Georgia, United States of America
\\
\textbf{3} Initiative in Theory and Modeling of Living Systems, Emory University, Atlanta, Georgia, United States of America

\bigskip

%
%





* catalina.maria.rivera@emory.edu

\end{flushleft}
\section*{Abstract}

Biochemical processes in cells are governed by complex networks of many chemical species interacting stochastically in diverse ways and on different time scales. Constructing microscopically accurate models of such networks is often infeasible. Instead, here we propose a systematic framework for building phenomenological models of such networks from experimental data, focusing on accurately approximating the time it takes to complete the process, the First Passage (FP) time. Our phenomenological models are mixtures of Gamma distributions, which have a natural biophysical interpretation. The complexity of the models is adapted automatically to account for the amount of available data and its temporal resolution. The framework can be used for predicting behavior of various FP systems under varying external conditions. To demonstrate the utility of the approach, we build models for the distribution of inter-spike intervals  of a morphologically complex neuron, a Purkinje cell, from experimental and simulated data. We demonstrate that the developed models can not only fit the data, but also make nontrivial predictions. We demonstrate that our coarse-grained models provide constraints on more mechanistically accurate models of the involved phenomena.

\section*{Author summary}
Building microscopically accurate models of biological processes that offer meaningful information about the behavior of these systems is a hard task that requires a lot of prior knowledge and experimental data that are not available most of the time. Here instead we propose a mathematical framework to infer  {\em phenomenological} models of biochemical systems, focusing on approximating the probability distribution of time it takes to complete the process. We  apply the method to study statistical properties of spiking in morphologically complex neurons, Purkinje cells, and make nontrivial predictions about this system.


\section*{Introduction}

Processes in living cells are governed by complex networks of stochastically interacting biochemical species. Understanding such processes holistically does not necessarily imply having a detailed description of the system at a microscopic, mechanistic level. Indeed, many microscopic networks can result in equivalent experimentally observable behaviors~\cite{belIlya2009}, so that distinguishing alternative networks may be impossible. Even if competing models are not exactly equivalent, they may approximate each other in many key measurable behaviors~\cite{gutenkunst2007universally}. Thus a lot of ink has been expended on developing methods for constructing reduced, coarse-grained models of biological processes as alternatives to unidentifiable mechanistically accurate ones~\cite{sinitsyn2009adiabatic, machta2013parameter, transtrum2015perspective,borisov2008domain,hlavacek2006rules,chylek2015modeling,conzelmann2008exact,munsky2006finite,haseltine2002approximate,kim2017reduction,kang2013separation,anderson2011model,huang2005systematic,rao2014model,maurya2005reduced}. This is usually a challenging task, requiring construction of a (possibly inaccurate) detailed mechanical model as an intermediate step. In this paper, we focus on an alternative approach of {\em refining} phenomenological models of stochastic biological processes rather than coarse-graining mechanistic ones. Our approach optimally adapts the level of complexity to match the amount and quality of the experimental observations while accurately predicting specific macroscopic properties of the processes.

A large number of biological processes -- and the sole focus of this work -- can be viewed as First Passage (FP), or completion processes~\cite{redner2001guide,iyer2016first,chou2014first,bressloff2013stochastic,zhang2016first}: certain molecules must interact, certain compounds must be created, or certain states must be visited, before an event of interest occurs.  For such systems, one is often interested in when the final event occurs (i.e., what the FP time is), rather than in  details of which molecules got created or which states were visited in the process.
Thus such systems represent a fruitful field for coarse-grained modeling.
Crucially, often the available experimental data are sufficiently precise to allow investigation of the whole probability distribution of the FP time, and the fact that the time is stochastic and often broadly distributed can have important functional effects~\cite{raj2008nature,iyer2016first,bressloff2014stochastic,munsky2012using}.

A natural approach to characterizing the FP distribution based strictly on the statistical information contained in the samples of the FP time involves progressively estimating its higher order cumulants.
However, this approach suffers from a well-known problem that such cumulant expansions cannot be truncated at any order but the second, and still give rise to a proper probability distribution~\cite{wallace1958asymptotic}.
Here we propose a different method for systematically inferring phenomenological models of first passage distributions from empirical data.
The approach does not strive for the mechanistic accuracy.
Instead, following ideas from~\cite{daniels2015}, we develop a family of models of FP processes, whose complexity can be grown adaptively as data requires, to fit arbitrary FP time distributions.
We then choose the optimal model of the appropriate complexity within the family using Bayesian model selection~\cite{schwarz1978estimating,kass1995bayes,chipman2001practical,rissanen1999hypothesis,balasubramanian1997statistical,mackay2003information}.

Our model family consists of mixtures of Gamma  distributions, which we argue to have a natural interpretation in the context of FP kinetics.
In the well-sampled regime, this natural interpretation allows us to infer mechanistic constraints on the underlying kinetics using fits within our model family~\cite{Kolomeisky}.
Specifically, the element of the mixture that dominates the passage for short times, sets the minimal number of internal states that a mechanistically accurate stochastic process would need to generate the data.
Furthermore, our approach provides a framework to study effects of external perturbations or experimental conditions on the first passage statistics in a systematic way. Specifically, by doing model selection simultaneously on all data sets across multiple experimental conditions, we can obtain a single phenomenological model that explains all of the available data, relating parameters of such global model to the values characterizing the perturbations.

We test the utility of our approach on neurophysiological data sets.
Most neurons are too complex to be modeled mechanistically with molecular accuracy, so that any model will involve some element of phenomenology, making this a good testing ground for our approach.
Indeed, spontaneous activity of neurons of different types is often modeled under the assumption that the spike trains can be described by renewal processes~\cite{tuckwell1988,Correia1977,Rodieck1962,Grossman1961,Lamarre1971,Steriade1973,Tolhurst1981}.
Since, in such models, all inter-spike intervals (ISIs) are independent and identically distributed, the spike generation can be specified fully by the ISI distribution, and hence can be seen as a FP process in our framework.
While one usually models the ISI distribution as a Gamma distribution~\cite{Bishop1964,Correia1977}, more complex constructions are often warranted~\cite{Burns1976,Tuckwell1978}.
Ability of our method to adapt the complexity of the model as required by the quality and the quantity of the data thus promises to be useful in this context.
To investigate this, we build models describing the ISI distribution in a certain type of neurons, called Purkinje cells (PCs), under a variety of experimental conditions, and with data coming from real experiments and from biophysically realistic models of the neuron.
Purkinje cells are some of the most morphologically  complex neurons, and, indeed, we discover that even {\em phenomenological} models of their ISI distributions need to be a lot more complex than a single Gamma distribution.
For example, we show that 5 or 6 terms in the mixture are needed to describe PCs of a Rhesus monkey.
At the same time, even the most detailed computational model of the process is fitted well with just 4 terms, hinting at a room for improvement of biophysical models. 

We conclude this article with a discussion of other applications where our method may be useful.

\section*{Results}
\label{Results}
\subsection*{The model family}
\label{TheModel}
\begin{figure}[H]
\begin{center}
\includegraphics[scale=0.5]{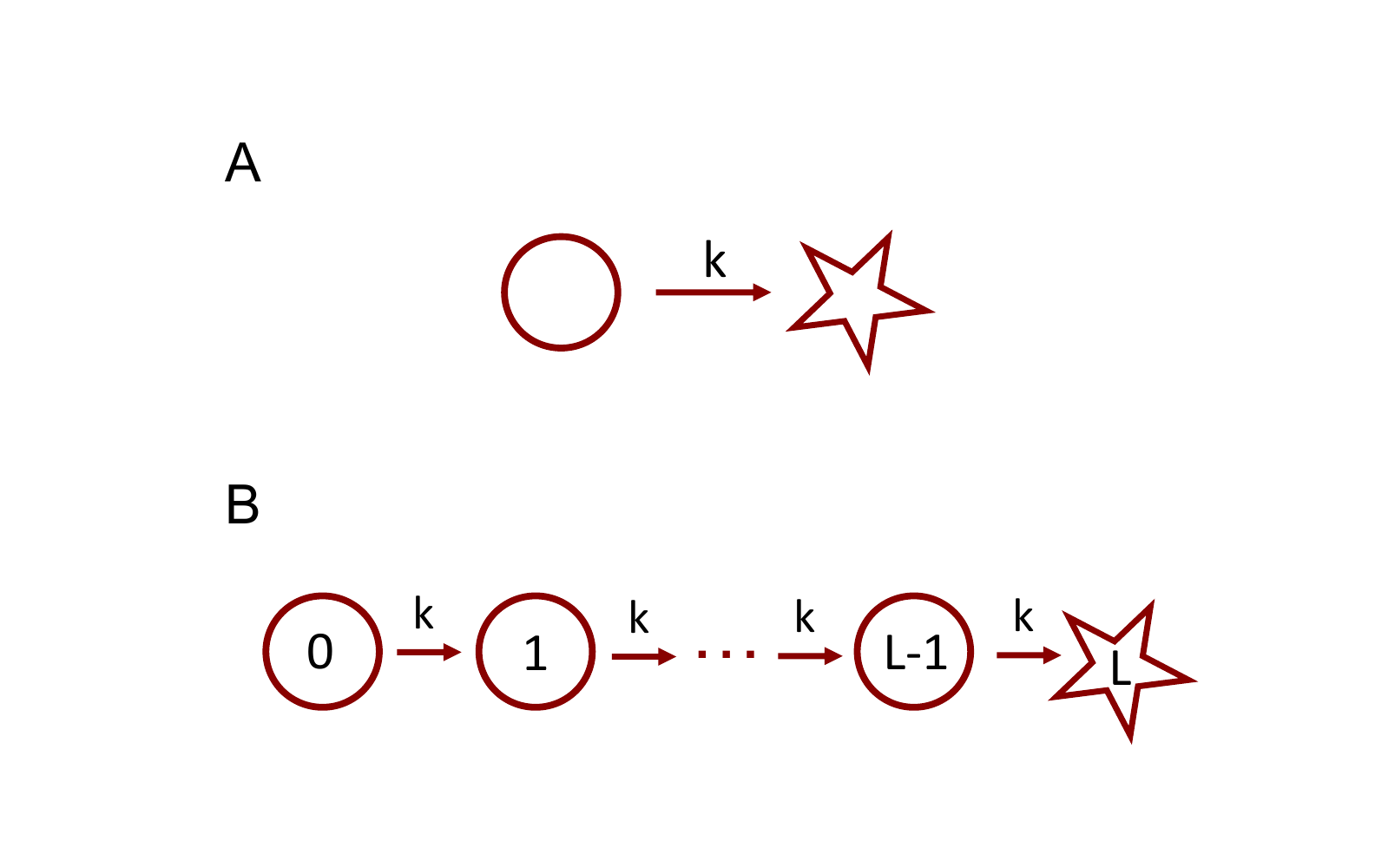}
\caption{\textbf{Simple FP processes.} A: Exponential completion, with $k=1/\tau$. B: Multi-step completion, with the Gamma-distributed completion time.}
\label{SimplestModel}
\end{center}
\end{figure}

\begin{figure}[H]
\begin{center}
\includegraphics[scale=0.4]{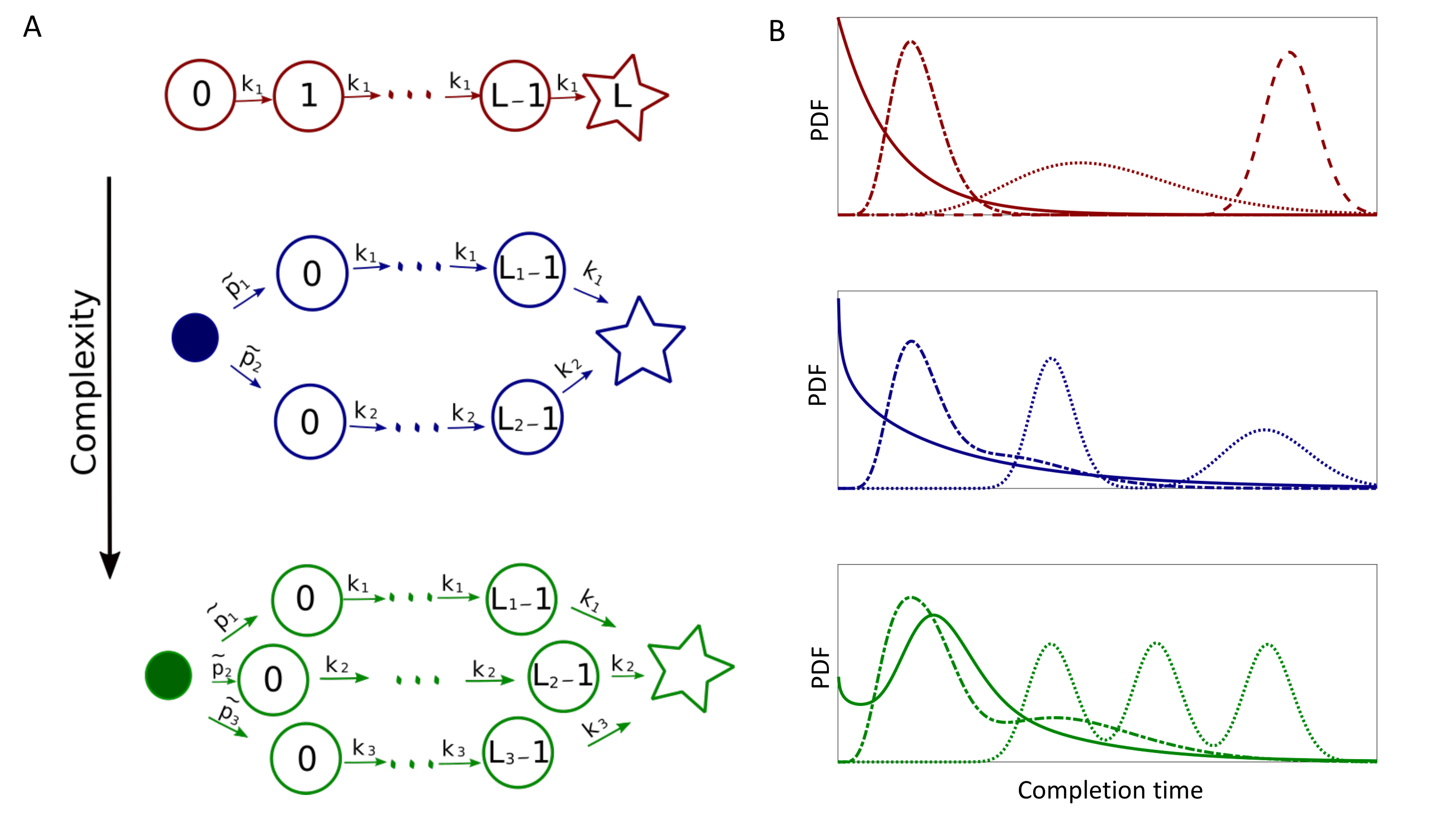}
\caption{\textbf{Hierarchical set of models.} A: Kinetic schemes of the first three models in the hierarchical set. Each next model in the hierarchy is built by adding another completion path, where $k_{i}=1/\tau_{i}$ is the transition rate between intermediate states, and ${p}_{i}$ is the probability of completion through the path $i$. B: Examples of FP probability densities that can be generated with the corresponding models with different parameter values.}
\label{ModelScheme}
\end{center}
\end{figure}

The simplest possible stochastic model to represent a FP process is a two state system as shown in Fig.~\ref{SimplestModel}A.
With a constant transition time $\tau$ between the initial and the absorbing state, we get an exponentially decaying completion time probability distribution $ P(t)= \exp(-t/\tau)/\tau$.
A natural extension is a multi-step activation process, where the system irreversibly passes through a number of intermediate states before reaching the absorbing state, see Fig.~\ref{SimplestModel}B.
A simple induction shows that the completion probability distribution in this case is given by the Gamma distribution, Eq.~(\ref{GammaDist2}):
\begin{equation}\label{GammaDist2}
  P(t|\tau,L)=\dfrac{t^{L-1}}{\tau^{L}(L-1)!}\exp(-t/\tau), 
\end{equation}
where $L$ corresponds to the number of intermediate states before FP and $\tau$ is the average transition time between the intermediate states, which we take to be the same for all states for simplicity and, as we show later, without the loss of generality.
This simple model is commonly used to describe neural ISI distributions.
However, often times  neural spikes  exhibit more complex ISI distributions~\cite{bair1994power,debusk1997stimulus,nowak2003electrophysiological,shih2011improved,tsubo2012power,sungho2016}.
Motivated by these empirical findings, we built a set of models that are hierarchically organized, so that their complexity can be adapted to the quality and the quantity of empirical data by adding additional Gamma-distributed completion paths as shown in Fig.~\ref{ModelScheme}A.

The mathematical expression of our model with $M$ different completion paths is:
\begin{equation}\label{GeneralModel}
\begin{aligned}
 P(t\mid \vec{\theta}, M)&=p_{1}P(t|\tau_{1},L_{1})+ p_{2}P(t|\tau_{2},L_{2})+... + p_{M}P(t|\tau_{M},L_{M}),\\
 p_{1}&=\frac{1}{1+x_{2}+...+x_{M}}, \quad p_{2}=\frac{x_{2}}{1+x_{2}+...+x_{M}},\quad ...\quad,\\ 
 p_{M}&=\frac{x_{M}}{1+x_{2}+...+x_{M}},
    \end{aligned}
\end{equation}
where $\vec{\theta}=( \tau_{1}, L_{1}; x_{2},\tau_{2},L_{2};\dots;x_{M},\tau_{M},L_{M})$ are parameters to be fitted and $P(t|\tau_{i},L_{i})$ are defined as in Eq.~(\ref{GammaDist2}).
Notice that when there is only one completion path, $M=1$, with only one non-absorbing state $L_{1}=1$, we recover the exponential distribution function with the decay time $\tau_{1}$.
Figure~\ref{ModelScheme}B shows examples of FP time distributions that can emerge from models with different small values of $M$ by changing parameter values.
These distributions can approximate processes, such as neuronal bursts, which have multiple characteristic time scales.

\setkeys{Gin}{draft=false}

  %

We will call the union of all models $P(t|\vec{\theta},M)$, with $M=1,\dots, \infty$, the {\em  multi-path} {\em model family} of FP distributions.
We will focus on Bayesian inference of phenomenological models of FP processes within this family for the rest of this work. 
One would like such statistical inference to be {\em consistent}, so that, in the limit of infinite data, one would recover the true model if it belongs to the model family being used in the inference.
For an infinite model family to allow such consistent statistical inference using Bayesian approaches, it is sufficient for the family to be {\em nested} and {\em complete}~\cite{Nemenman2005}.
Nestedness (or hierarchy) means that models within the family can be ordered in such a way that the set of solutions of a given model is contained in the set of solutions of the next model in the hierarchy.
Completeness means that every data set can be fitted arbitrarily well by some (possibly very complex) model in the hierarchy. 

The multi-path model family is trivially nested: if we set $p_M=0$, then the model with $M$ paths reduces to the one with $M-1$.
The proof of completeness is a bit more subtle, see {\em Methods}.
With that, we know that estimating the posterior probability of the model within the family given the observed data $D$, and then choosing $M$ that maximizes the posterior probability $P(M\mid D)$, will typically result in consistent inference and in ``selection'' of the most probable model.
Specifically, we need to evaluate
\begin{equation}\label{BayesianModelSel}
 P(M\mid D) \propto \int P(D\mid \vec{\theta}, M) P(\vec{\theta} \mid M) d\vec{\theta},
\end{equation}
where
\begin{equation}\label{likelihood}
 P(D\mid \vec{\theta}, M) = \prod_i P_M(t_i|\vec{\theta}),
\end{equation}
and $t_i$ is the $i$'th completion time in the experimental data set being fitted.
Evaluating this integral and hence building the most probable phenomenological model of the data is the goal of our paper.

Unfortunately, as $M$ grows in Eq.~(\ref{BayesianModelSel}), the involved integral becomes high-dimensional, and it is very difficult to estimate reliably.
One usually assumes that the integrand is strongly peaked near the {\em maximum likelihood} value $\vec{\theta}_0$, which maximizes $P(D\mid \vec{\theta}, M)$.
A variety of approximate methods exist for the evaluation~\cite{schwarz1978estimating,tierney1986accurate,geweke1989bayesian,smith1993bayesian,kass1995bayes,ding2018model}, which make different assumptions about the structure of the integrand near its maximum likelihood argument $\vec{\theta}_0$.
We observed that, for most data sets we tried, $P(D\mid \vec{\theta}, M)$ were far from Gaussian, thus prohibiting the use of the simple Laplace approximation to evaluate the integral~\cite{schwarz1978estimating,tierney1986accurate}.
Therefore,  we used importance sampling~\cite{Owen2013,geweke1989bayesian} to evaluate  Eq.~(\ref{BayesianModelSel}), see {\em Methods}.

Experimental data is usually quantized in units of the experimental time resolution.
To fit such data we, therefore, transform Eq.~(\ref{GeneralModel}) into its discrete time version by integrating FP probabilities over a time discretization window $\Delta t$.
That is, Eq.~(\ref{GeneralModel}) becomes 
\begin{align}\label{DiscreteGeneralModel}
 P_{\Delta t}(t\mid \vec{\theta}, M)=&p_{1}\int_{t-\Delta t}^{t}P(t|\tau_{1},L_{1})dt\\\nonumber&+ p_{2}\int_{t-\Delta t}^{t}P(t|\tau_{2},L_{2})dt+\dots + p_{M}\int_{t-\Delta t}^{t}P(t|\tau_{M},L_{M})dt\\\nonumber
 \approx &p_{1}P(t|\tau_{1},L_{1})\Delta t+ p_{2}P(t|\tau_{2},L_{2})\Delta t+\dots + p_{M}P(t|\tau_{M},L_{M})\Delta t.
\end{align}

The code to implement the {\em multi-path model family} for FPP is available at https://github.com/criver9/Inferring-FPP.git

\subsection*{Model for interspike intervals for Purkinje cells}
\label{Results for PC}

Purkinje Cells (PCs) are neurons present in the cerebellum of  vertebrate animals, which participate in learning.
They have large and intricate dendritic arbors and produce complex action potentials with a multiscale distribution of the interspike intervals (ISIs).
Due to the complexity of the cells, their typical models involve many dozens of compartments, each described by a handful of biophysical parameters~\cite{DeSchutter1994, Miyasho2000, santamaria2002modulatory, kulagina2007electro, forrest2012sodium}.
Crucially, the process of generating a spike can be seen as a FP process, where the neuron goes through a set of different effective states, not necessarily in a simple sequence, before crossing a certain voltage threshold (the absorbing state that results in a spike generation).
Thus here we ask whether the ISI distribution for PCs, indeed, requires so many features to model well, or if, in contrast, the structural complexity of PCs does not result in a similarly high  complexity of the spike generation.
To answer this, we use ISIs of PCs corresponding to simple spikes (spontaneously generated by the cell) of a Rhesus monkey ({\em Macaca mulatta}), obtained from~\cite{sungho2016}, and we search for the best phenomenological model of this distribution using our approach. 

\begin{figure}[H]
\begin{center}
\hspace*{-1.5cm}\includegraphics[scale=0.55]{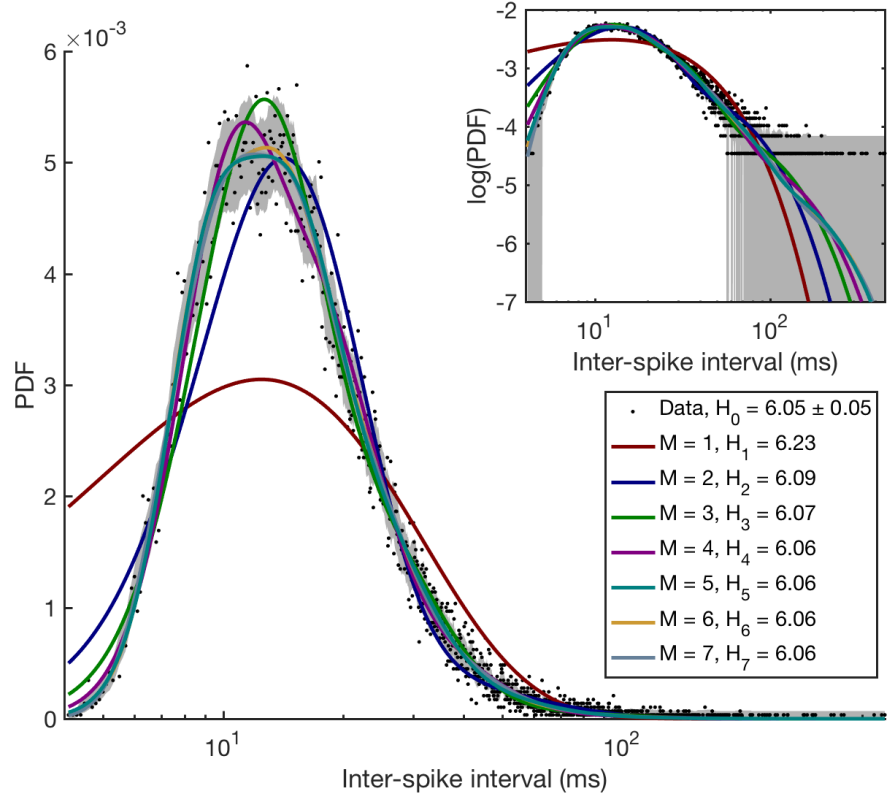}
\caption{\textbf{Best fit models, $M=1\dots 7$, for Purkinje cells  ISI distribution.} Dots indicate the histogram of the real data, and the grey band denotes the standard error of every dot. Color lines show the average fit line sampled from the posterior distribution of each of the first seven models in the hierarchy, error bands (too narrow to see) on these fits where estimated using the standard deviation from the sampled curves, see {\em Methods}.  The legend illustrates how the cross-entropy between the data and the models decreases with the model complexity towards the data entropy. Note that the horizontal axis is logarithmic. Inset: same data, but on log-log axes. }
\label{RealPCFit}
\end{center}
\end{figure}

Figure~\ref{RealPCFit} shows the best fits for each of the model in our hierarchy, $M\le 7$, to the PC ISI distribution data. 
The figure and Tbl.~\ref{ProbModelPurk} suggest that
the simplest phenomenological model of the process contains  about $M=5$ effective independent paths (for this data set, we cannot discriminate between models with $M=5,6$ based on the values of $P(M\mid D)$).
Notice that, by  gradually adding additional completion paths, we can approximate not only the right tail of the distribution, but also the left tail --  the behavior at early times.
We measure this quality of fit by showing, in  Fig.~\ref{RealPCFit}, the entropy, $H_0=-\sum_{i=1}^{N}p_{i} \ln p_{i}$ (evaluated using the NSB estimator~\cite{nemenman2002entropy}), of  data being fitted, as well as the cross-entropy entropy, $H_M=-\sum_{i=1}^{N}p_{i} \ln P_{\Delta t}(t_{i}\mid \vec{\theta}, M)$, between the data and each of the best fit models with different $M$ (this corresponds to minus the normalized value of the log-likelihood, Eq.~(\ref{likelihood})).
To the extent that $H_M$ approaches $H_0$ for larger $M$, the fits are quite good.
And since $H_M\approx H_{M+1}$ for large $M$, the fits stop becoming much better, so that the Bayesian Model Selection~\cite{mackay2003information} then penalizes models with large $M$, forcing us to settle at $M\approx 5$.

We next check how the selected model depends on the amount of data being fitted.
As seen in Tbl.~\ref{TableVsN}, increasing the number of spikes in the data set from 1000 to $\sim 30000$ allows us to identify finer details in the data which require more accurate models to be fitted.
Thus the most likely model has $M=2$ for a small data set, gradually increasing to $M=5$ for full data.
Since the last three-fold increase in the amount of data does not result in a further growth of the best $M$, we conclude that the phenomenological model likely has reached the complexity needed to explain the system, and the model with $M=5$ is, in some sense, equivalent to the full complexity of simple spike generation of a real Purkinje cell.

This analysis illustrates two crucial points.
First, a relatively simple model with $M\approx 5$ is able to explain the {\em experimental} ISI distribution from a complex neuron, so that much of the physiological complexity of the cell {\em does not} translate into a functional complexity, at least at the scale of a simple spike generation.
Second,  {\em quantitatively} fitting the data favors models with $M\ge 5$ by a factor of $\sim 10^{20}$ (see Tbl.~\ref{ProbModelPurk}).
In other words, PC spiking is not trivially simple, and guessing this ISI model without the automated inference procedure developed here  would likely be impossible.

\begin{table}[H]
\centering
\begin{tabular}{ |p{0.5cm}|p{2.5cm}| }
\hline
$M$&$\ln P(D\mid M)$ \\
\hline
1 &-180379 \\
2 &-176368 \\
3 &-175826\\
4 &-175694\\
\rowcolor{lightgray}5 &\textbf{-175649}\\
6 &-175651\\
7 &-175653\\
\hline
\end{tabular}
\caption{\textbf{Model selection results for ISI of experimental PC.} Posterior likelihoods of the first seven models in the model family are shown for $N=28966$ spikes (the full data set). The model with the highest marginal likelihood, $M=5$, is highlighted. Note that models with $M=6,7$ cannot be ruled out, as they have very similar posterior likelhoods.}
\label{ProbModelPurk}
\end{table}

\begin{table}[H]
\centering
\begin{tabular}{ |p{0.5cm}|p{1.8cm}|p{1.8cm}|p{1.8cm}|p{1.8cm}|p{1.8cm}| p{1.8cm}| }
\hline
\multicolumn{7}{|c|}{$\ln P(D\mid M)$} \\
\hline
M& 1000 &5000 & 10000 & 15000 & 20000 & 28966 \\
\hline
1 &-6192.08 &-30901 & -61792         &-92841         & -124040 & -180379\\
2 &\textbf{-6133.59}           &-30427        &-60673 &-91189 & -121736 & -176368\\
3 &-6133.88          &-30369         &-60536          &-90915          & -121356 & -175826\\
4 &-6142.02           &\textbf{-30349}        &-60486        &-90858           & -121262 & -175694\\
5 &-6151.81           &-30350         &\textbf{-60481}         &\textbf{-90840}           & \textbf{-121230} &\textbf{-175649}\\
6 & ----------- &-30357 & -60485& -90844&-121232 &-175651\\
\hline
\end{tabular}
\caption{\textbf{Model selection as a function of the number of samples.} First row shows the size of the data set, $1000\dots 28966$, and the rest of the table shows the posterior probability of each model in the family for these data. As the number of samples increases, more complex models are required to explain the details of PC spiking, but the complexity eventually saturates, presumably having matched the complexity of the real cells  observed at the given experimental accuracy. }
\label{TableVsN}
\end{table}

\subsection*{Model for ISI of synthetic PC}
\label{ResultsSyntheticPC}
One of our interests is to develop phenomenological models that are able to predict the change in the FP distributions for a system under the influence of various external perturbations.
We would like to illustrate this using PCs.
However, we are not aware of readily available large, precise data sets measuring the ISI distribution in PCs under external perturbations.
Thus instead we focus on synthetic data, generated using a biophysically realistic, multi-compartmental model that resembles the morphologically complex structure of PCs, the  Miyasho et al.\ model~\cite{Miyasho2000}, which is a  modified version of the earlier De Schutter and Bower model~\cite{DeSchutter1994}.
To illustrate the complexity of the  Miyasho model, we point out that it uses 1087 compartments to describe the dendritic arbor of a PC and one compartment for the soma.
Additionally, the dynamics are defined by around 150 parameters that specify 12 different types of voltage-gated ion channels~\cite{Miyasho2000}. 

\begin{figure}[H]
\begin{center}
\hspace*{-3.5cm}\includegraphics[scale=0.4]{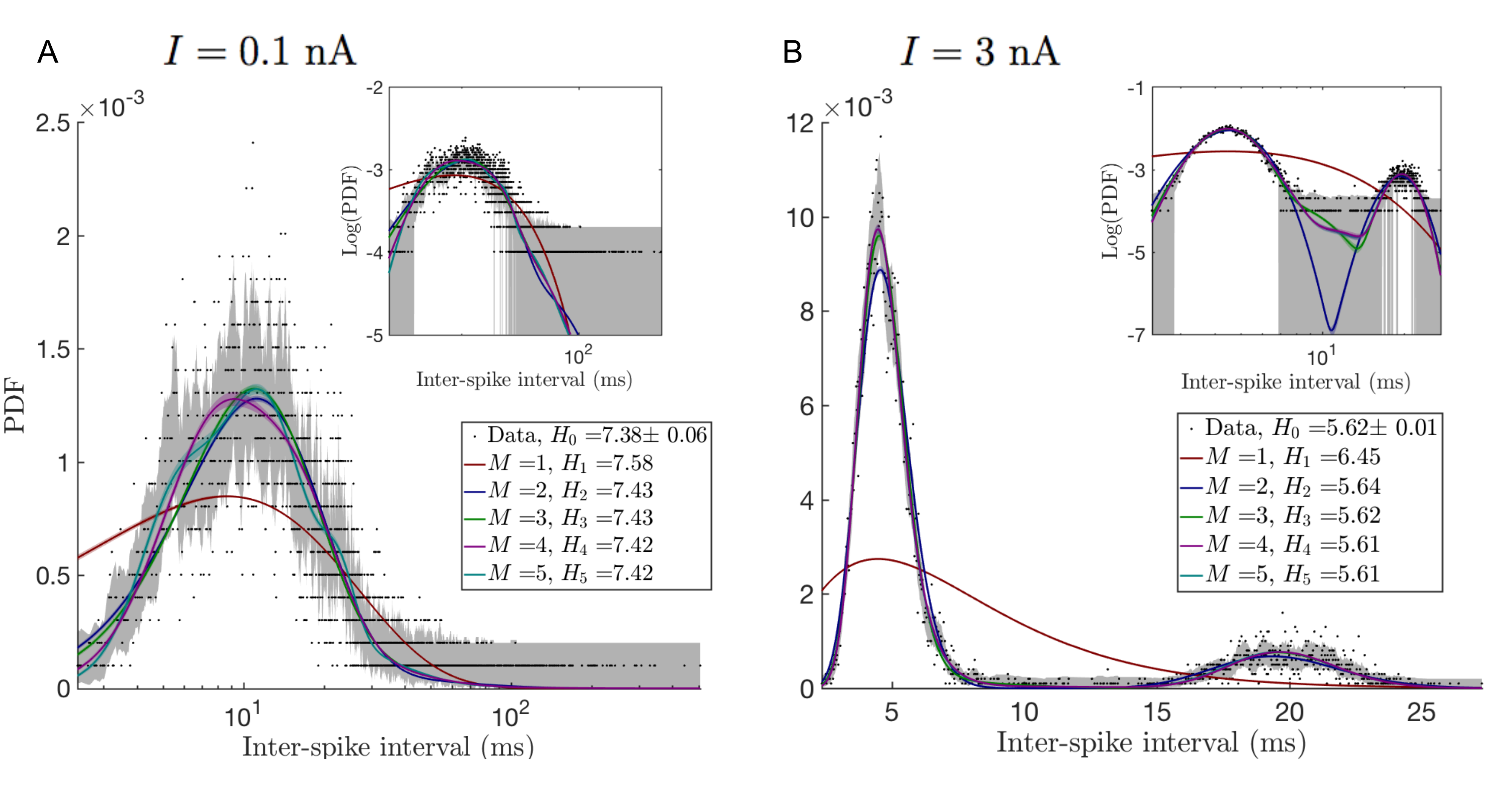}
\caption{\textbf{Best fits for different models in the model family for the distribution of ISIs of  synthetic PCs.} Color lines and color bands show the mean and standard deviation respectively of different models sampled from the posterior distribution of each of the first five models in the family (see details in the Appendix~\ref{S3_AppendixBestCurves}).  The legend shows how the cross entropy decreases with the model complexity towards its minimum value of the entropy of the histogram of the observed data. According to Tbl.~\ref{ProbModelPurk2}, 4 paths are needed  to explain the ISI characteristics of synthetic PCs under different external conditions. A: injected current $I=0.1$ nA, and B: $I=3$ nA.}
\label{SynthPCFit}
\end{center}
\end{figure}

We used this model to simulate the behavior of the membrane potential dynamics of a PC, affected by different electric currents injected into the soma.
White noise currents with standard deviation $\sigma=3$ nA and mean values ${I}=0.1, 0.5, 0.7, 1, 2, 3$ nA where injected, thus generating six different data sets, with which to explore the ISI probability distributions of the PC model.
Following the procedure described earlier, we selected the simplest phenomenological model that can explain the ISI statistics of the PC model, but in this case we focus on optimizing the posterior likelihood over {\em all} stimulus values  simultaneously.
Figure~\ref{SynthPCFit} shows the best model fits for two different injected currents which produce qualitatively different  ISI distributions.
Fits for other current values can be found in Fig.~\ref{S1_Fig}.
To build the optimal model for all injected currents simultaneously, we estimate the posterior likelihood of each model in the family for $M\le 5$ for each of the synthetic data sets, see Tbl.~\ref{ProbModelPurk2}.
Since for different currents, the ISI generated are independent of each other, the log-likelihood for the entire data set is simply the sum of log-likelihoods for each $I$.
As always, we choose the optimal model as the one with the largest overall log-likelihood.

\begin{table}[H]
\begin{adjustwidth}{-1.25in}{0in}
\centering
\begin{tabular}{ |p{0.5cm}|p{1.8cm}|p{1.8cm}|p{1.8cm}|p{1.8cm}|p{1.8cm}|p{1.8cm}| p{1.8cm} |}
\hline
\multicolumn{8}{|c|}{$ \ln P(D\mid M)$} \\
\hline
$M$& $I=0.1$ nA &$I=0.5$ nA & $I=0.7$ nA & $I=1.0$ nA & $I=2.0$ nA & $I=3.0$ nA & Total \\
\hline
1 &-75437 &-71282 &-66821 & -66309&-64654 &-64488&-408992\\
2 &-74070 &-70034 &-65283 &-63578 &-58932 &-56462&-388359\\
3 &-74019 &-70001 &\textbf{-65238} &-63520 &\textbf{-58773} &\textbf{-56211}& -387762\\
\rowcolor{lightgray}4    &-74003 &-69993 & -65239&\textbf{-63516} &-58794 &-56212& \textbf{-387753}\\
5 &-73994&-69976&-65249&-63527*&-58805*&-56223*&-387775\\
\hline
\end{tabular}
\caption{\textbf{Model selection results for ISI of synthetic PCs.} Posterior likelihood of the first five models in the family for each data set, corresponding to the six different injected currents. Last column shows that a model with $4$ completion paths is optimal  over the combined data. Asterisk marks those cases where the optimal parameter values fell at the boundary of the search space, usually because there were paths with near-zero flux through them (see {\em Methods}). Note that the numbers in the first two columns increase monotonically with $M$, so that the best model in the family is not found for $M\le 5$. We chose to truncate the exploration at $M=5$ since we are interested in the overall maximum of the log-likelihood for all $I$, which is reached at $M=4$ (last column).}
\label{ProbModelPurk2}
\end{adjustwidth}
\end{table}

\begin{figure}[H]
\begin{center}
\hspace*{-5cm}\includegraphics[scale=0.35,angle=0,origin=c]{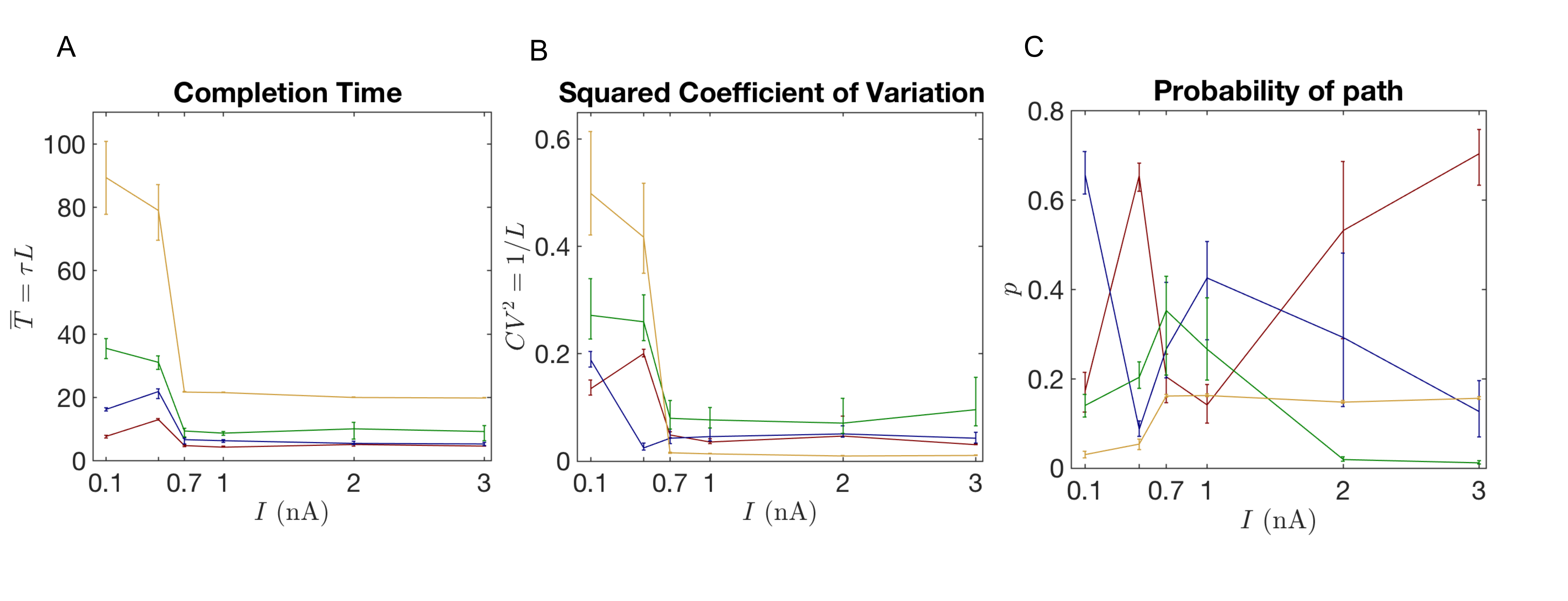}
\caption{\textbf{Properties of completion paths change as a function of the external parameter for the best model selected across all experiments.} A: Average completion times for each of the $M=4$ independent paths are plotted as a function of the injected current in the soma, $I$. Color (same in (B) and (C)) identifies paths according to how long they take to complete the process on average. B: Coefficient of variation and C: probability of taking each of the independent paths of the model as a function of $I$.}
\label{SynthParamVsI}
\end{center}
\end{figure}

Table~\ref{ProbModelPurk2} shows that, for our data sets, $M=4$ effective independent paths are enough to explain simultaneously the PCs behavior under six different injection currents. As can be seen in Fig.~\ref{S2_Fig}, when the injected current increases the cell goes from the non-bursting to the bursting state, and the entropy of the completion time distribution decreases (see Figs.~\ref{SynthPCFit},~\ref{S1_Fig}). Table~\ref{ProbModelPurk2} indicates that higher entropy distributions, corresponding to $I=0.1, 0.5$ nA need $M \geq 5$ completion paths to be properly explained. Lower entropy distributions, on the other hand, not only require fewer paths, but also more deterministic paths, as can be observed from the coefficient of variation estimates in Fig.~\ref{SynthParamVsI}. This suggests, that under low external stimulus ($I < 0.5$ nA), spike generation in the cell can happen through multiple pathways. Instead, when a certain current threshold is reached ($I > 0.5$ nA), only a few of these pathways get activated. Nonetheless, more than one pathway is needed even for high currents, since, at least, two time scales are required to explain the bursting activity.

In Fig.~\ref{SynthParamVsI}, we explore how the properties of the model selected in Tbl.~\ref{ProbModelPurk2} ($M=4$) change as a function of the injected current, $I$. Each independent path is described by specifying its average completion time $\overline{T}_{i}=\tau_{i} L_{i}$, the coefficient of variation ${\rm CV}^2_{i}=1/L_{i}$, and the probability $p_{i}$ of completion along this path, and these three quantities are plotted for each path for different values of $I$. There is a sharp change in these features when the  PC transitions from a non-bursting to a bursting state, between $I=0.5$ and $0.7$ nA. For example, completion times and coefficient of variation for all paths drop drastically at this point. In particular, Fig.~\ref{S3_Fig} shows that the paths with the longest completion time explain very different aspects of the non-bursting and the bursting ISI distributions. For the non-bursting cases, these paths help to fit mostly the tails. Instead, for the bursting cases, these paths explain the intra-burst time interval, which happens to be a much more deterministic process, as can be seen from the behavior of the coefficients of variation, Fig.~\ref{SynthParamVsI}.

\begin{figure}[H]
\begin{center}
\hspace*{-4cm}\includegraphics[scale=0.53,angle=0,origin=c]{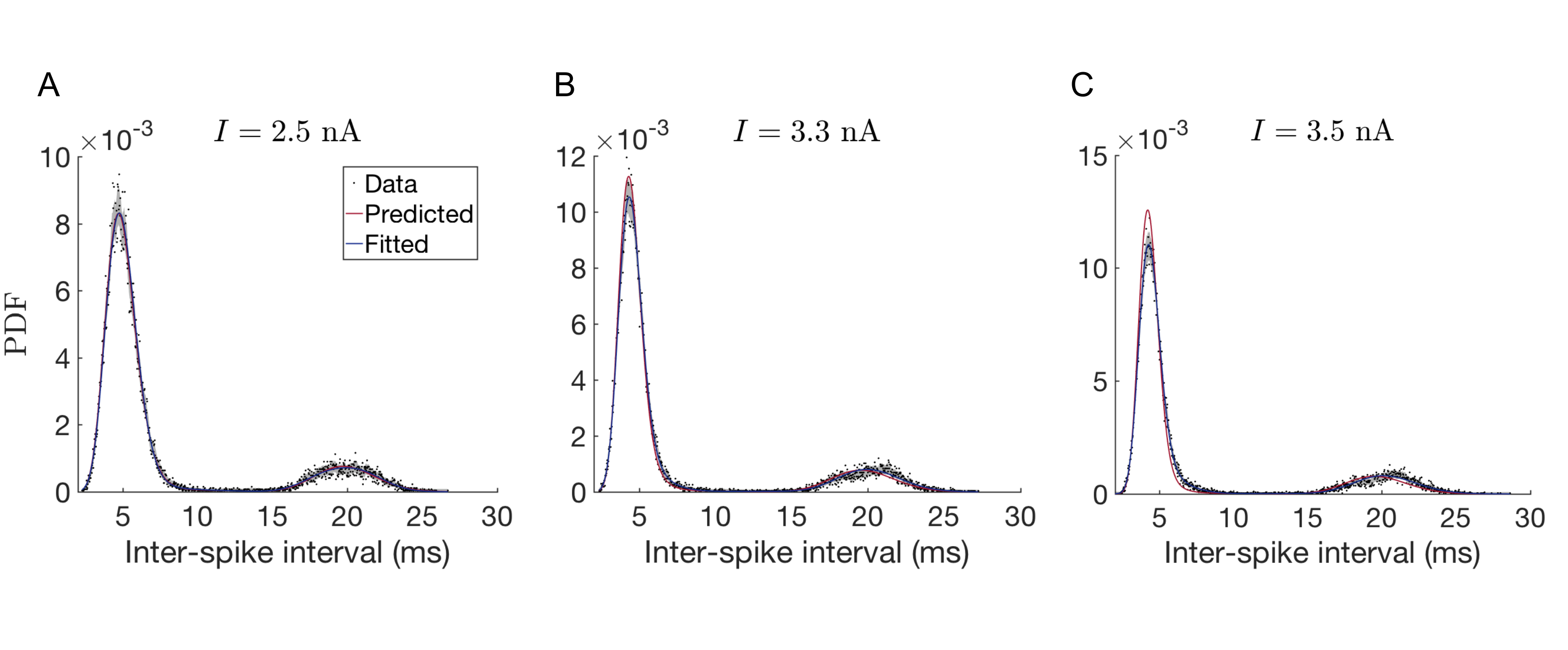}
\caption{\textbf{Predicted PDFs} for non-measured values of the injected current. Predicted model (in red) was obtained by interpolating parameter values from Fig.~\ref{SynthParamVsI}. It is compared with the model (in blue) fitted directly to data. (A) Prediction for $I=2.5$ nA (interpolation). (B) and (C) Prediction for $I= 3.3$ and $I =3.5$ nA respectively (extrapolation).}
\label{Predictions}
\end{center}
\end{figure}

To test whether the phenomenological model correctly captures the time scales of the underlying biophysical processes, we predict the ISI distribution for input currents that the model was not exposed to during fitting.
To achieve this we first need to determine a relationship between model parameters and the input current means, which we can then use to infer model parameters for currents different from the ones used for fitting the model.
As our test case, we employed the model with $M=4$ and tracked the dependence of its parameters on the current as shown in Fig.~\ref{SynthParamVsI}.
A priori it is unclear how to build correspondence between the four model paths for separate input currents.
In our example in Fig.~\ref{SynthParamVsI}, we chose to establish the correspondence by ordering the paths according to their completion time, thus relating the model paths with the smallest completion time, then the second to smallest and so on.
This ordering provides relationships between input currents and all model parameters, based on which we can infer parameter values for new current values using linear interpolation (for currents that fall between two fitted values) or linear extrapolation (for currents outside of the fitted range).
We note that the choice to relate parameter values by completion time rather than another parameter is arbitrary.
Indeed there are many possibilities to create the pathway correspondence for different current values.
Besides ordering based on average completion time (confront Fig.~\ref{SynthParamVsI}) we also tested ordering based on the coefficient of variation or the probability path which led to no improvement over the presented case (not shown).
While it is possible that other orderings can lead to better predictions we leave a more systematic exploration of this aspect for future work.

To validate our predictions, we generated new data for mean currents $I=2.5$, $3.3$, and $3.5$ nA and compared predicted ISI distributions to the simulation results (see Fig.~\ref{Predictions}).
The predicted model for $I=2.5$ nA (where parameters interpolate between the known values at $I=2.0$ and $3.0$ nA) is almost indistinguishable from the fitted one (Fig.~\ref{Predictions}A).
Even the extrapolation to  $I=3.3$  and $I=3.5$ nA (Fig.~\ref{Predictions}B-C) show very good agreement between predicted model and simulation data.

\begin{figure}[H]
\begin{center}
\hspace*{-4cm}\includegraphics[scale=0.4]{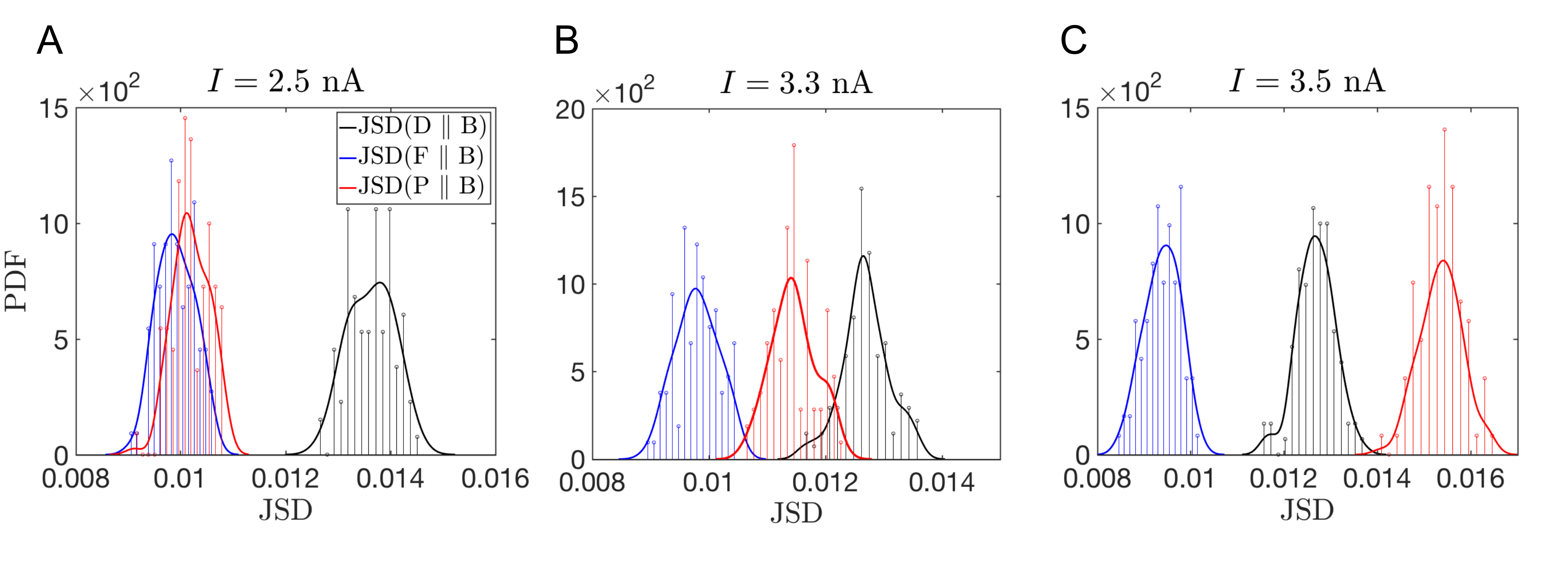}
\caption{{\bf Quantifying quality of the predictions.} We plot the histograms of the JSD between the test data set and the bootstrapped samples from the validation data set (in green), the  JSD between the bootstrapped validation data sets and models fitted to each of these data sets (in blue), and the JSD between the bootstrapped data and the prediction based on interpolating or extrapolating the model parameters fitted to the original data (in red). To the extent that the distributions are close, predictions are good. A-C: Data for $I=2.7, 3.3,  3.5$ nA, respectively. The first is interpolation, the other two are extrapolations.}
\label{FigS6}
\end{center}
\end{figure}

To quantify the accuracy of these predictions, we need to calculate their quality with respect to some baseline.
We chose the Jensen-Shannon Divergence (JSD)\cite{lin1991} as a measure of the quality of fit, and we measure it relative to two baselines.
First, we quantify how an extrapolated or an interpolated prediction compares to the fit done directly on a data set; certainly the fit is expected to outperform the prediction.
Second, we check how two statistically equivalent realizations of data fit each other; this should be the ceiling, which neither the fit nor the prediction can outperform (if both are not overfitted).
Both of these baselines depend on the specific data set used, and thus one needs to estimate probability distributions of the relevant JSDs, rather than their single values.
However, generating data from the PC model takes hours even on a modern computer, and hence we generate only a single additional, validation, data set beyond the training and the testing sets, which we then additionally bootstrap (resample with replacement) to produce statistics of the JSDs.
Specifically, Fig.~\ref{FigS6} plot histograms of (i) the JSD between the test data and the bootstrapped versions of the validation data (this is the statistics that requires us to have two independent samples, test and validation, to remain unbiased), (ii) the JSD between the bootstrapped validation data and fits to these data, and (iii) the JSD between the prediction and the bootstrapped validation data.
Our first observation is that  all three JSD distributions are very close to each other, indicating very good fits and predictions.
For $I=2.5$ nA, the fits/predictions have smaller JSD than different realizations of data have with themselves, which is consistent a very good fit, and suggests, as expected, that the variability across bootstrapped data sets is somewhat larger than would have been across independent samples.
As $I$ increases, and interpolation gives way to extrapolation, the prediction quality deteriorates (still remaining only a few percent worse than the fits).

\subsubsection*{Inferring mechanistic constraints}
\label{MinConst}
Our approach to modeling FP time probability distribution is purely phenomenological.
However, the {\em  multi-path} {\em model family} allows us additionally to constrain mechanistic, biophysical models of the underlying processes.
Specifically, we can make predictions for the minimal number of intermediate states that a mechanistic model requires to explain the data.
Indeed, for any FP problem, the short-time behavior of the completion probability density provides information about the length of the shortest completion path~\cite{Kolomeisky, valleriani2014unveiling}.
That is, assume that the process starts in a state $i$ and ends at the absorbing state $j$ of an arbitrary Markovian chemical reaction network.
Then, at short times, the completion probability density can be approximated as $\rho_{ij}\propto t^{m}$, where $m$ is the number of intermediate states of the shortest path connecting states $i$ and $j$~\cite{valleriani2014unveiling}.
In principle, this means that by estimating the exponent of the power law that fits the left tail of the completion time distribution, one can put a lower limit on the number of intermediate states in a mechanistic model.
Then any candidate model with a fewer number of steps can be rejected.

In practice, making use of this result is hard because it requires data with very high temporal resolution, and a very well sampled left tail.
However, our {\em  multi-path} representation allows for an extension of the approach to the case where the sampling is good, but the time resolution may not be sufficient for simpler methods.
Once the most probable model in the model family is selected and fitted, we propose to determine if the first few fastest events can be explained by a single independent path $i$ of length $L_{i}$.
We use 50 events in our analysis, which provides for a sufficient number of the events to seek a power law fit, and yet is small enough so that only the very end of the left tail is explored.
Since at short time scales the Cumulative Distribution Function (CDF) of the FP time probability density is $\propto t^{L_i}$ (from Eq.~(\ref{GammaDist2} )), one can insist that any mechanistic model built to describe the data will need at least $L_{i}$ states, establishing a lower bound on the size of the network.

\begin{figure}[H]
\begin{center}
\hspace*{-5cm}\includegraphics[scale=0.47]{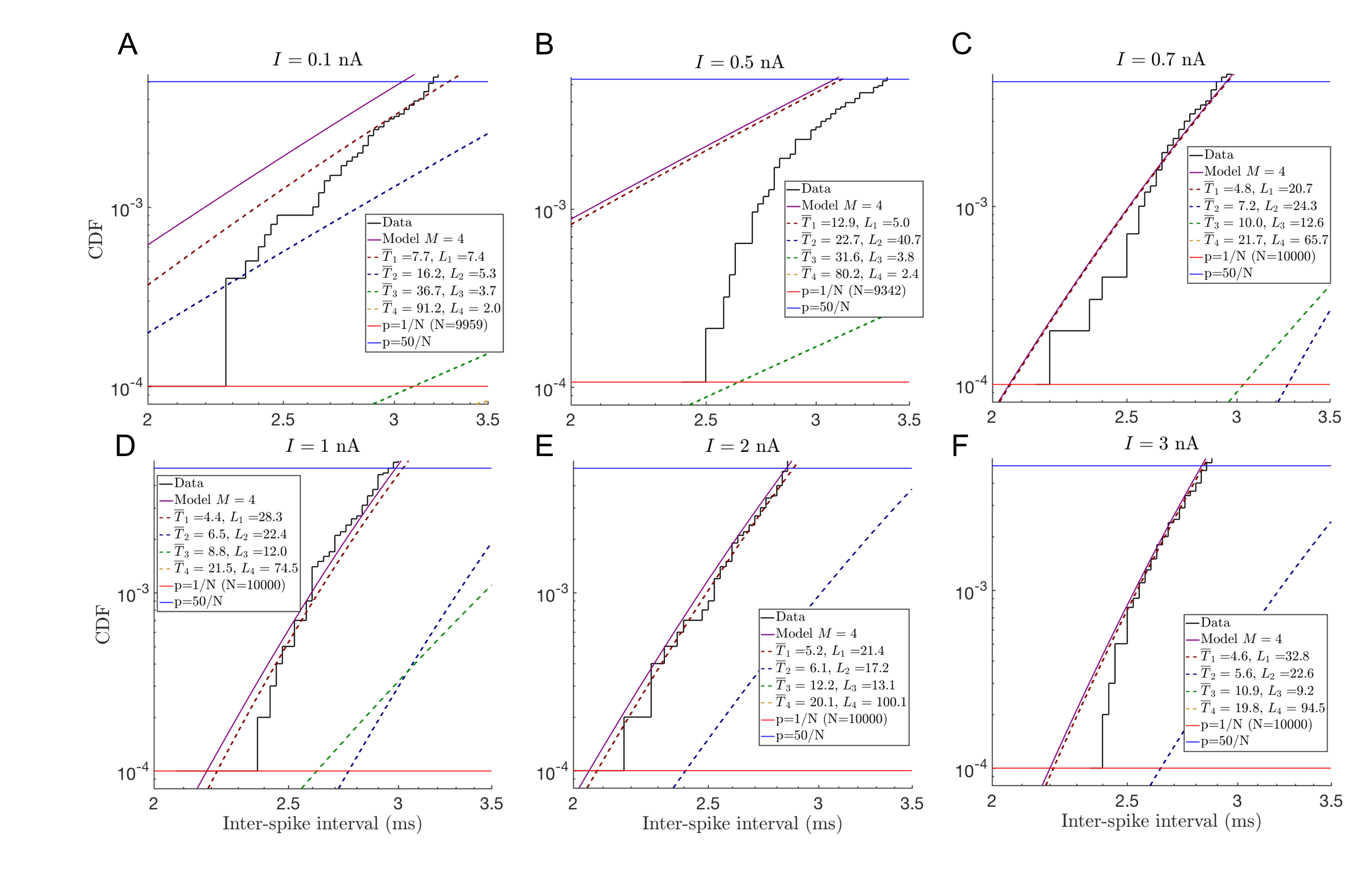}
\caption{\textbf{Decomposition of the Cumulative Distribution Functions (CDFs) of completion time  at early times into the four completion pathways.} Black line represents the CDFs from data; horizontal red and blue lines in each plot correspond to the probability of the 1st and the 50th events, respectively. Solid purple lines are CDFs from the best-fitted model with $M=4$, and each of the dashed lines represents contributions from the constituent completion paths. A-F: $I=0.1, 0.5, 0.7, 1.0, 2.0, 3.0$ nA, respectively. Panels C-F show that the first fifty events can be explained by one or more  paths with $\sim 20\dots30$ intermediate states. Therefore, any biophysically accurate reaction network explaining these data   needs to have at least $>20$ internal states. Notice that, even though a model with $M=4$ is optimal over all values of $I$ according to  Tbl.~\ref{ProbModelPurk2}, it does not explain the early time behavior in panels (A,B). }
\label{S5V_Fig}
\end{center}
\end{figure}

For concreteness, the short time behaviors of the CDFs obtained from the best model, $M=4$, describing the ISIs of PCs under six different injected currents are shown in Fig.~\ref{S5V_Fig}. Only for $I=0.7$ nA the first 50 events ($0.5\%$ of sample size) can be explained by a single path with $ \sim 20$ intermediate states, while for larger values of $I$, the distribution can be fitted by one or more of such paths.
In all of these cases, it is thus clear that any realistic biophysical model of a PC must include,  at least, $\sim20$ internal states.

\section*{Discussion}

In this study, we developed a mathematical structure of the ({\em  multi-path} {\em model family}) to infer phenomenological models describing FP time distribution for biological processes. As an example of application of our approach, we show that this representation allows us to build models capable of describing the complexity of the ISI distributions of PCs by successfully explaining not only the bulk, but also the tails of the distribution. Our results show that the process of a spike generation in PCs is more complex than a simple renewal process with a Gamma-distributed completion time, which is typically used in the field. For spontaneously generated spikes,  $M\geq5$ independent Gamma-distributed paths are required. We also showed that only $M \approx 4$ paths (11 independent parameters) are needed to explain the behavior of synthetic PCs over all injected current values $I> 0.5$ nA. This illustrates that (i) morphological complexity of PCs notwithstanding, their dynamics is not very complex at the level of the FP time distribution, and (ii) our fully phenomenological approach can, nonetheless, point out when biophysically-realistic models are inconsistent with features of experimental data. By identifying how parameters of the inferred model change with the external stimulus and extrapolating or interpolating them, we can predict the FP time distribution of the system in response to novel stimulus values. These predictions focus not just on the mean and the variance, but on the entire completion time distribution, and we have shown that the predictions are remarkably accurate, as compared to statistical fluctuations in the data themselves. Finally, we showed how our purely phenomenological approach can  establish the minimum size of a mechanistically accurate biochemical  network underlying the system, at least for  well-sampled data sets.

The specific model family hierarchy we developed here is only one of many possible hierarchies that are both complete and nested. Like in~\cite{daniels2015}, different hierarchies may be better suited for phenomenological modeling of different biological processes, and thus their relative success would reveal salient properties of the modeled processes. We hope to develop such additional hierarchies, and explore their pros and cons in the subsequent paper. Additionally, here we assumed that every completion time is independent and identically distributed. This is a strong assumption, which is not always realized. Even for PCs, the burstiness of spiking suggests dependence among the successive ISIs (i.e., within a burst, a short ISI is usually followed by another short ISI). In the future, it should be possible to extend our approach to model such processes by either modeling the statistics of FP time for a sequence of events, or by extending the model family to incorporate a latent variable that controls the dependence among subsequent completion events. 

Our  models offer only limited understanding of the mechanistic details of the modeled biological process.  Nonetheless, there are many advantages to our approach, and phenomenological modeling in general.  Indeed, the complexity that biological processes have acquired over eons of evolution oftentimes makes building detailed microscopic models an extremely challenging task. And yet the functional properties of the behavior might be rather simple, with the structural complexity existing, for example, to ensure robustness of the function to various perturbations. Then focusing on the phenomenological model allows us to elucidate, predict, and eventually use properties of the functional behavior even if microscopic details of the mechanisms used to produce it remain unclear. Our specific approach to phenomenological modeling is different from many others in that it does not coarse-grain a microscopic model (requiring a laborious task of building one as an intermediate step), but rather it  {\em refines} phenomenological models, adding progressively more details until the functional behavior is well approximated. Bayesian model selection is used to find the optimal point in the refinement hierarchy.  The computational advantages of taking such an adaptive, refining  approach can be huge, especially when the studied complex system exhibits a simple behavior. The computational complexity of our approach is dominated by searching for optimal fits, which scales linearly with the data set size, and exponentially with the model complexity. However, the latter is rarely more than a few dozen parameters even for very complex systems, such as the PCs, at least for realistic experimental resolution and data set sizes. Thus we expect our approach to be useful for modeling any biological system for which (i) the quantity that we need to predict is the completion time, (ii) the underlying biophysics is very complex, with microscopic details not always affecting the macroscopic completion properties, and where  (iii) large, high quality experimental data sets are available for different experimental conditions, requiring (iv) to predict  the behavior of the system as a function of these conditions, for their yet-untested values.

\section*{Materials and Methods}
\label{Methods}
\subsection*{Completeness}
\label{S1_AppendixModel}

Here we show that the model family studied in this work, Eq.~(\ref{DiscreteGeneralModel}), is complete.
That is, any data set describing the distribution of the completion times of the first passage process can be approximated arbitrarily well by a gamma mixture model with sufficient complexity.

We note that experimentally measured and numerically simulated completion times are constrained by finite resolutions which essentially discretizes the time axis.
Thus we can write the completion time likelihood as a multinomial 
\begin{equation}\label{Completeness1}
 L(\vec{q}\mid \vec{n})= q_{1}^{n_{1}}q_{2}^{n_{2}}...(1-q_{1}-q_{2}-...q_{K-1})^{(N-n_{1}-n_{2}-...n_{K-1})},
\end{equation}
where $n_{i}$ counts how often the completion time falls into the $i$th out of $K$ time interval bins $(t_{i}-\Delta t,t_{i}]$, $N$ is the total number of completion time events, and $q_{i}$ is the probability of completion in the time interval defined by bin $i$, given by $q_{i}=P_{\Delta t}(t_{i}\mid \vec{\theta}, M)$ (see Eq.~(\ref{DiscreteGeneralModel})).
Trivially, the maximum of $L(\vec{q}\mid \vec{n})$ is achieved when $q_{1}=n_{1}/N$, $q_{2}=n_{2}/N$, ...$q_{K}=n_{K}/N$.
Therefore, our aim must be to construct a model that can bring $\vec{q}$ arbitrarily close to this maximum.
The rational of the proof is to have a path per time bin whose average waiting time is the center of the respective time bin and whose variance can get arbitrarily small, effectively approximating a delta function.
That is, we want to construct a model such that for any $\epsilon>0$, we have $n_{i}/N-\epsilon \leq P_{\Delta t}(t_{i}\mid \vec{\theta}, K) \leq n_{i}/N + \epsilon$.

To prove this we set the parameters in Eq.~(\ref{DiscreteGeneralModel}) to what follows.
For the probability of every gamma path take $p_{i}=n_{i}/N$, with expected completion time given by $T_{i}=L_{i}\tau_{i}=t_{i}-\Delta t/2$ and variance (arbitrarily small) $\sigma_{i}^{2}=T_{i}\tau_{i}=\frac{\Delta t^{2}}{4}\epsilon_{i}$, where $\epsilon_{i}=\mathrm{min}(\frac{\epsilon}{p_{1}+p_{2}+...+p_{i-1}+p_{i+1}+...p_{K}}, \frac{\epsilon}{p_{i}})$.
Then, we can show that: 
\begin{equation}\label{DiscreteGeneralModelCompleteness}
 \begin{aligned}
 P_{\Delta t}(t_{i}\mid \vec{\theta}, K)&=\frac{n_{1}}{N}\int_{t_{i}-\Delta t}^{t_{i}}P(\tau_{1},L_{1})dt+ \frac{n_{2}}{N}\int_{t_{i}-\Delta t}^{t_{i}}P(\tau_{2},L_{2})dt+... + \frac{n_{K}}{N}\int_{t_{i}-\Delta t}^{t_{i}}P(\tau_{k},L_{K})dt\\
 &\leq \epsilon_{i}[\frac{n_{1}}{N}+...\frac{n_{i-1}}{N}+\frac{n_{i+1}}{N}+...\frac{n_{K}}{N}] +\frac{n_{i}}{N}\\
 &\leq \epsilon + \frac{n_{i}}{N}
 \end{aligned}
\end{equation}
where we used Chebyshev's inequality ($\mathrm{Pr}(|t-T_{i}| \geq \alpha \sigma_{i})\leq \frac{1}{\alpha^{2}}$, with $\alpha=1/\sqrt{\epsilon_{i}}$) to set a bound to all the integrals but the $i$th.
For the $i$th integral we note that, since most of the probability mass falls in this bin, it reaches close to one and is naturally bounded by one.
This concludes the upper bound on the $q_i$.
For the lower bound we simply subtract one from both sides of the Chebyshev inequality and multiply by negative one to get $\mathrm{Pr}(|t-T_{i}| \leq \alpha \sigma_{i})\geq 1-\frac{1}{\alpha^{2}}$.
This gives a bound for the $i$th integral of Eq.~(\ref{DiscreteGeneralModelCompleteness}):
\begin{equation}\label{UpperBoundCompleteness}
 \begin{aligned}
 P_{\Delta t}(t_{i}\mid \vec{\theta}, K)\geq  \frac{n_{i}}{N}\int_{t_{i}-\Delta t}^{t_{i}}P(t|\tau_{i},L_{i})dt
 &\geq \frac{n_{i}}{N} (1-\epsilon_{i})\geq \frac{n_{i}}{N}-\epsilon,
 \end{aligned}
\end{equation}
showing that this model family can approximate any sufficiently smooth distribution arbitrarily well.
In real applications, we may not need to have as many paths as there are bins to achieve a high approximation accuracy, so the construction above is the worst case scenario.

\subsection*{Model Selection}
\label{S2_AppendixBayesianSel}
 
To choose the most likely model from the family, we evaluate and maximize the posterior probability of each model $M$:
\begin{equation}\label{BayesSel}
 P(M\mid D)=\frac{P(D \mid M)P(M)}{P(D)} \propto P(D\mid M),
\end{equation}
where we assumed that all models in the hierarchy are a priori equally likely.
The likelihood $P(D \mid M)$ is given by:
\begin{equation}\label{BayesSel2}
 P(D\mid M)= \int P(D\mid \vec{\theta}, M) P(\vec{\theta} \mid M) d\vec{\theta},
\end{equation}
where the likelihood of the data set and the prior are chosen to be:
\begin{align}
  P(D\mid \vec{\theta}, M) &= \prod_{i=1}^{K}P_{\Delta t}(t_{i}\mid \vec{\theta}, M)^{n_i}, \label{Likelihood2}\\
  P(\vec{\theta} \mid M) &= \frac{1}{(Z_{x})^{M-1}}\prod_{j=1}^M
  \frac{\exp(-\frac{\tau_{j}}{Z_{\tau}})}{Z_{\tau}} \frac{\exp(-\frac{L_{j}}{Z_{L}})}{Z_{L}}.\label{Likelihood3}
\end{align}

Here $P_{\Delta t}(t\mid \vec{\theta}, M)$ is given by Eq.~(\ref{DiscreteGeneralModel}), and $n_i$ is the number of events with completion time between $(t_{i}-\Delta t, t_{i})$.
The parameters of our prior are $Z_{L}$, $Z_{\tau}$, and $Z_x$.
The values of $Z_{L}$ and $Z_{\tau}$ are set such that the priors are reasonably wide compare to the measured time scales and throughout our study we set them to $Z_{L}=20$ and $Z_{\tau}=20$ ms.
$Z_x$ sets the upper boundary for the support of the $x_i$ and was set to $Z_{x}=10^{3}$.
Finally, we note that our choice of prior assumes no correlation among model parameters.

In most cases, the integration in  Eq.~(\ref{BayesSel2}) is analytically intractable.
A typical approach in such a case is to use the Laplace approximation to compute the integral.
However, in our considered problems the posterior distributions fall much slower than Gaussians, ruining the quality of the Laplace approximation.
Thus we used importance sampling~\cite{Owen2013} instead.
Specifically, we sampled from the multi-variate normal distribution $G(\vec{\theta})= \det(2\pi \Sigma)^{-\frac{1}{2}}\exp{(-\frac{1}{2}(\vec{\theta}-\vec{\theta}^{*})'\Sigma^{-1}(\vec{\theta}-\vec{\theta}^{*}))}$ centered at the optimal value $\vec{\theta}^{*}$ of the integrand $ F(\vec{\theta}):=P(D\mid \vec{\theta}, M) P(\vec{\theta} \mid M)$ with the covariance matrix $\Sigma$ defined by the Hessian of $F(\vec{\theta})$:
\begin{equation}\label{HessianDef}
 (\Sigma^{-1})_{ij}= (-\Hessian \log F\mid_{\vec{\theta}^{*}})_{ij}  \equiv - \frac{\partial^{2} \log F}{\partial \theta_{i} \partial \theta_{j} }\Big\rvert _{\vec{\theta}^{*}}.
\end{equation}
This way we ensured that $G(\vec{\theta})>0$ for $F(\vec{\theta})>0$, at least around the domain of the local optimum at $\vec{\theta}^{*}$. 
See below for details of how we estimated the covariance matrices.
Then the importance sampling estimate of the integral in Eq.~(\ref{BayesSel2}) is
\begin{equation}\label{ImportSamp}
 P(D\mid M) \sim \frac{1}{N} \sum_{i=1}^{N} \frac{P(D\mid \vec{\theta}_{i}, M) P(\vec{\theta}_{i} \mid M)}{G(\vec{\theta}_{i})},
\end{equation}
where $\vec{\theta}_{i} \sim \mathcal{N}(\vec{\theta}^{*},\Sigma)$ and we used $N=10^{6}$ samples to achieve the desired accuracy.
Since the likelihood values exceeded numerical resolution, we instead computed the $\ln P(D \mid M)$:

\begin{equation}\label{ImportSamp2}
 \ln P(D\mid M) \sim  \ln F(\vec{\theta}^{*}) + \ln   (\sum_{i=1}^{N}\frac{\exp(\log F(\vec{\theta}_{i}) - \log F(\vec{\theta}^{*}))}{G(\vec{\theta}_{i})})-\ln N.
\end{equation}

\subsubsection*{Covariance matrix estimation}
\label{CovEstimation}

Application of our importance sampling scheme requires knowing the maximum of the integrand and the Hessian around the optimum.
The optimal values $\vec{\theta}^{*}$ were obtained using the MATLAB function {\tt fminsearchbnd}.
We used MATLAB version R2017a for our analysis.
Most of the optimal values obtained for different models and data sets fell in the interior of the parameter's domain; we mark those where the optimal values fell at the boundary with an asterisk everywhere in the text. 

We first explain how we computed the covariance matrix for the cases where the optimal values fell in the interior of the parameters' domain set.
Using Eq.~(\ref{Likelihood2}) to estimate the Hessian, we get
\begin{equation}\label{HessianDefSI}
\begin{aligned}
 - \frac{\partial^{2} \log F}{\partial \theta_{k} \partial \theta_{j} }\Big\rvert _{\vec{\theta}^{*}}&=- \sum_{i}^{M} n_i \frac{\partial^{2} \log (P_{\Delta t}(t_{i}\mid \vec{\theta}, M))}{\partial \theta_{k} \partial \theta_{j} }\Big\rvert _{\vec{\theta}^{*}}- \frac{\partial^{2} \log P(\vec{\theta} \mid M)}{\partial \theta_{k} \partial \theta_{j} }\Big\rvert _{\vec{\theta}^{*}}\\
 &= \sum_{i}^{M} n_i \bigg[ \frac{1}{P_{\Delta t}(t_{i}\mid \vec{\theta}, M)^{2}}\frac{\partial P_{\Delta t}(t_{i}\mid \vec{\theta}, M)}{\partial \theta_{k}}\frac{\partial P_{\Delta t}(t_{i}\mid \vec{\theta}, K)}{\partial \theta_{j}}\Big\rvert _{\vec{\theta}^{*}}\\
 &-\frac{1}{P_{\Delta t}(t_{i}\mid \vec{\theta}, M)}\frac{\partial^{2}P_{\Delta t}(t_{i}\mid \vec{\theta}, M)}{\partial \theta_{k} \partial \theta_{j}}\Big\rvert _{\vec{\theta}^{*}}\bigg].
\end{aligned}
\end{equation}
Notice that the contribution to the Hessian coming from the prior in the previous expression cancels out. We then evaluated Eq.~\ref{HessianDefSI} numerically using Eq.~(\ref{DiscreteGeneralModel}).

For those cases, for which the optimal values are located at the boundary of the parameters' domain due to the presence of a trivial completion path we use the following trick. Given that the flux through a certain path $j$ is zero, the likelihood $P(D\mid \theta,M)$ stays constant for all values of $\tau_{j}$ and $L_{j}$ corresponding to this trivial path. However, the prior decays exponentially  and therefore $F(\vec{\theta})$ also decays exponentially in the directions of $\tau_{j}$ and $L_{j}$. The  optimal value of $F(\theta)$ can be written as $(\vec{x}_{p},x_{d}=0, \tau_{d}=0, L_{d}=0 )$ with $\vec{x}_{p}$ is the best fit for the previous model in the family, with only $d-1$ completion paths. Then the covariance matrix is:
\begin{equation}\label{HessianEdge}
\Scale[2]{\Sigma=}\left(\begin{array}{@{}c|c@{}}
  \begin{matrix}
   \Scale[2]{\Sigma_{p}}
  \end{matrix}
  & \bigzero\\
\hline
  \bigzero &
  \begin{matrix}
  \alpha_{x}^{2} & 0 & 0 \\
  0 & \alpha_{\tau}^{2} & 0 \\
  0 & 0 & \alpha_{L}^{2}
  \end{matrix}
\end{array}\right),
\end{equation}
where $\Sigma_{p}$ is the covariance matrix at the best fit of the previous model in the family; $\alpha_{x}^{2}$ is an upper bound on the variance along the parameter controlling the probability flux through $d$-th completion path estimated from the symmetric function $F_{s}(\theta)=F(|\theta|)$. We used $\alpha_{x}=0.01$ for all the cases marked with an asterisk in Tbl.~\ref{ProbModelPurk2}. On the other hand $\alpha_{\tau}$ and $\alpha_{L}$ where estimated using the variances of the independent exponential distributions of the prior, Eq.~(\ref{Likelihood3}), $Z_{\tau }=Z_{L}=20$. We chose $\alpha_{\tau}^{2}=\alpha_{L}^{2}=(3\sigma_{\tau})^{2}=3600$. Notice that, along these last two directions where $F(\theta)$ decays exponentially  we chose the variance of the importance distribution nine times larger in these two directions to make sure that it contains most of the important domain of $F(\theta)$.

\subsubsection*{Parameter Degeneracy}

The posterior distributions that we obtain often have multiple modes that correspond to parameter degeneracy, which arises by relabeling the completion paths. To account for this degeneracy in caclulating posterior likelihoods, we multiplied the likelihoods of each model with $M$ gamma pathways by $(M-1)!$. Here we use $M-1$ instead of $M$ because the first is different from the others: transition rate to this path is set to one and  is used as a references.

\subsubsection*{Generalized Bayesian model selection}

In order to find the model in the family that best fits the simultaneous description of the system under $s$ different external conditions, we need to estimate the integral Eq.~(\ref{BayesSel2}) for $s$ independent data sets,
\begin{equation}\label{GeneralizedSelection}
\begin{aligned}
 P(D_{1},D_{2},...,D_{s}\mid M)&= \int P(D_{1},D_{2},...,D_{s}\mid \vec{\theta}_{1},\vec{\theta}_{2},...,\vec{\theta}_{s}, M) P(\vec{\theta}_{1},\vec{\theta}_{2},...,\vec{\theta}_{s} \mid M) d\vec{\theta}\\
 &= \prod_{j=1}^{s}\int P(D_{j}\mid \vec{\theta}_{j}, M) P(\vec{\theta}_{j} \mid M) d\vec{\theta_{j}}.
\end{aligned}
\end{equation}
The last equality results from each data set having its own, independent set of parameters.
Taking the natural logarithm on both sides of Eq.~(\ref{GeneralizedSelection}), we obtain the following result, which we used to compute the values in Tbl.~\ref{ProbModelPurk2}:
\begin{equation}\label{GeneralizedSelection2}
 \ln P(D_{1},D_{2},...,D_{s}\mid M)= \sum_{j=1}^{s}\ln P(D_{j}\mid M)
 \end{equation}

\subsection*{Expected values and uncertainty of fits}
\label{S3_AppendixBestCurves}

The fits and the error bars for curves for all of the fitted models in all Figures are the expected values and the standard deviations of the model curves over the posterior probability distributions. That is,
\begin{eqnarray}\label{MeanCurve}
  \langle f(t\mid M)\rangle &=& \int f(t \mid \vec{\theta}, M)P(\vec{\theta} \mid D , M)d\vec{\theta},\\
\label{VarianceCurve}
 \Var(f(t\mid M))&=& \int (f(t \mid \vec{\theta}, M)-\left <f(t\mid M)\right >)^{2}P(\vec{\theta} \mid D , M)d\vec{\theta},
\end{eqnarray}
where $f(t \mid \vec{\theta},M)=P_{\Delta t}(t\mid \vec{\theta}, M)$, and the posterior probability is
\begin{equation}\label{Posterior}
 P(\vec{\theta} \mid D , M)=\frac{P(D\mid \vec{\theta},M)P(\vec{\theta}\mid M)}{P(D\mid M)}=\frac{F(\vec{\theta})}{P(D\mid M)}.
\end{equation}
As explained above, we used importance sampling to estimate the expectation values.
For example,  notice that Eq.~(\ref{MeanCurve}) can be rewritten as
\begin{equation}\label{MeanCurve2}
 \left <f(t\mid M)\right >= \frac{\int f(t \mid \vec{\theta}, M)F(\vec{\theta})d\vec{\theta}}{\int F(\vec{\theta})d\vec{\theta}}.
\end{equation}
Using Eq.~(\ref{ImportSamp2}), this becomes 
\begin{equation}\label{MeanCurve3}
 \left <\hat{f}(t\mid M) \right >\approx \frac{\sum_{i=1}^{N}\frac{f(t \mid \vec{\theta}_{i}, M)\exp(\log F(\vec{\theta}_{i}) - \log F(\vec{\theta}^{*}))}{G(\vec{\theta}_{i})}}{\sum_{i=1}^{N}\frac{\exp(\log F(\vec{\theta}_{i}) - \log F(\vec{\theta}^{*}))}{G(\vec{\theta}_{i})}}
\end{equation}

Similarly, for the variance, we have
\begin{equation}\label{VarianceCurve2}
 \Var(f(t\mid M))= \frac{\int (f(t \mid \vec{\theta}, M)-\left <f(t\mid M)\right >)^{2} F(\vec{\theta}) d\vec{\theta}}{\int F(\vec{\theta}) d\vec{\theta}},
\end{equation}
which results in
\begin{equation}\label{VarianceCurve3}
 \Var(\hat{f}(t\mid M))\approx \frac{\sum_{i=1}^{N}\frac{f(t \mid \vec{\theta}_{i}, M)^{2}\exp(\log F(\vec{\theta}_{i}) - \log F(\vec{\theta}^{*}))}{G(\vec{\theta}_{i})}}{\sum_{i=1}^{N}\frac{\exp(\log F(\vec{\theta}_{i}) - \log F(\vec{\theta}^{*}))}{G(\vec{\theta}_{i})}}- \langle \hat{f}(t\mid M)\rangle^{2}.
\end{equation}

\section*{Acknowledgments}
We thank Damian Hernandez, Baohua Zhou, Alejandro Rivera and Gordon Berman for the great ideas and discussions. This work was partially supported by the James S.\ McDonnell foundation Grant No.~220020321, and by the National Science Foundation Grants No.~1410978, 1806833, 1822677.

\section*{Supplementary Figures}
\label{SI}


\renewcommand\thefigure{S\arabic{figure}}    
\setcounter{figure}{0}

\begin{figure}[H]
\begin{center}
\hspace*{-4cm}\includegraphics[scale=0.45]{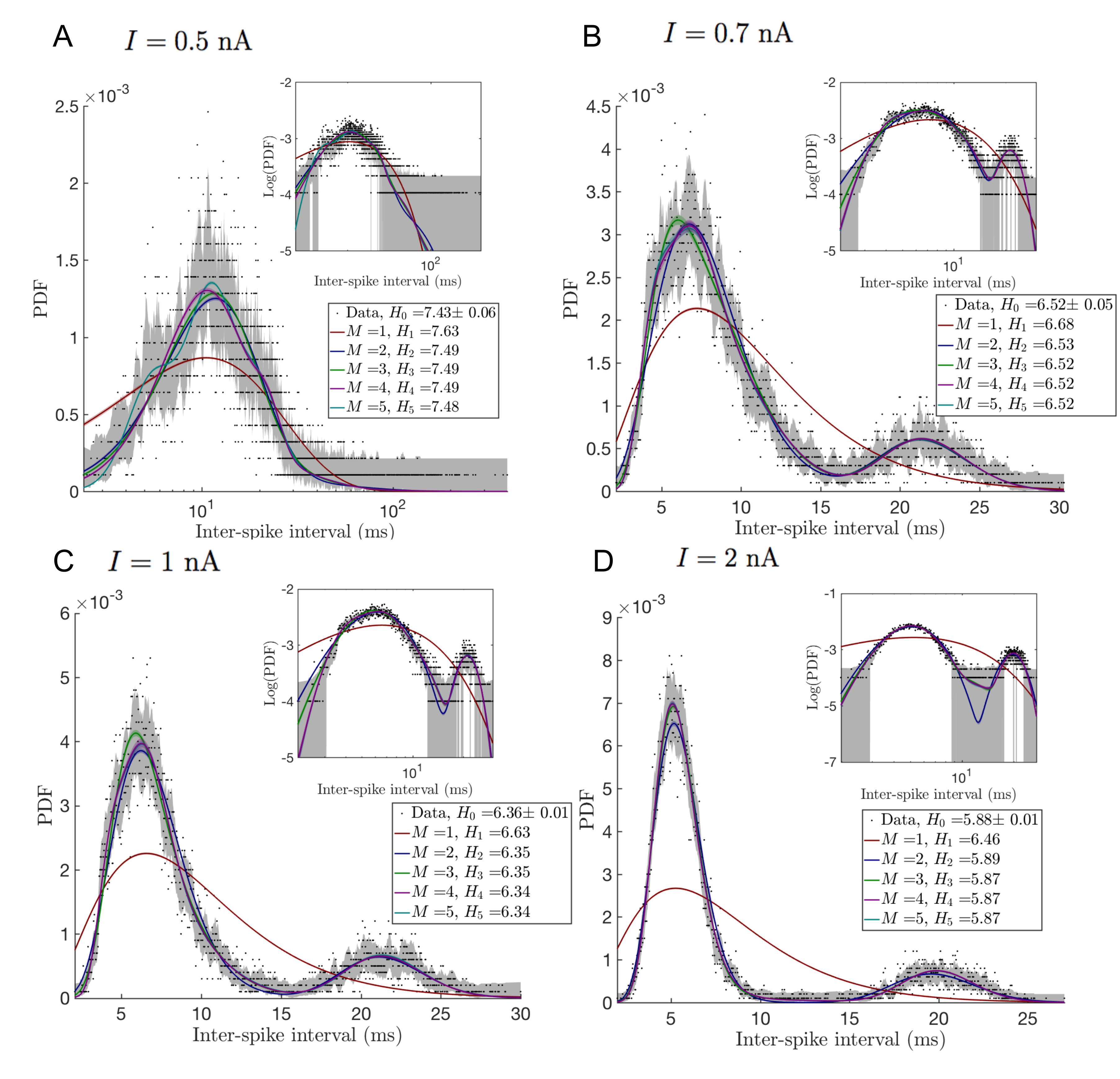}
\caption{\textbf{Best fits for different models in the family to the experimental Purkinje Cells Interspike Interval data.} Color lines and bands (the latter often too narrow to be seen) show the mean and the standard deviation of different models sampled from the posterior distribution of each of the first five models in the family.  The legends illustrate the decrease of the cross entropy with the model complexity towards its minimum value of the entropy of the histogram of the observed data. According to Tbl.~\ref{ProbModelPurk2}, 4 paths are needed to explain the ISI characteristics of synthetic under different external conditions. (A, B, C, D) injected currents $I=0.5, 0.7, 1.0, 2.0$ nA, respectively. }
\label{S1_Fig}
\end{center}
\end{figure}

\begin{figure}[H]
\begin{center}
\hspace*{-3cm}\includegraphics[scale=0.55]{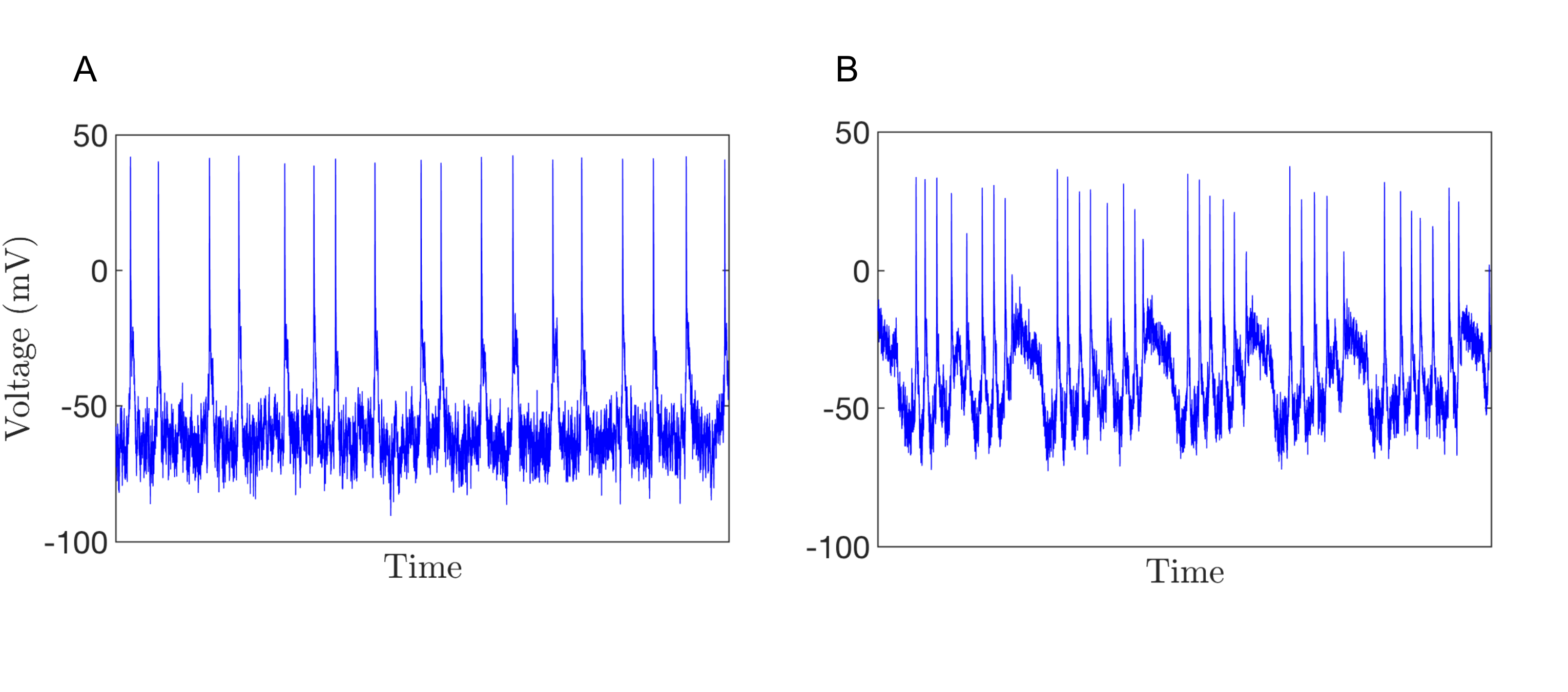}
\caption{\textbf{Simulated PC membrane potential} using the multi-compartmental model proposed in \cite{Miyasho2000} for A: low ($I=0.5$ nA) and B: high ($I=3$ nA) values of the injected current.  }
\label{S2_Fig}
\end{center}
\end{figure}
\begin{figure}[H]
\begin{center}
\hspace*{-3cm}\includegraphics[scale=0.5]{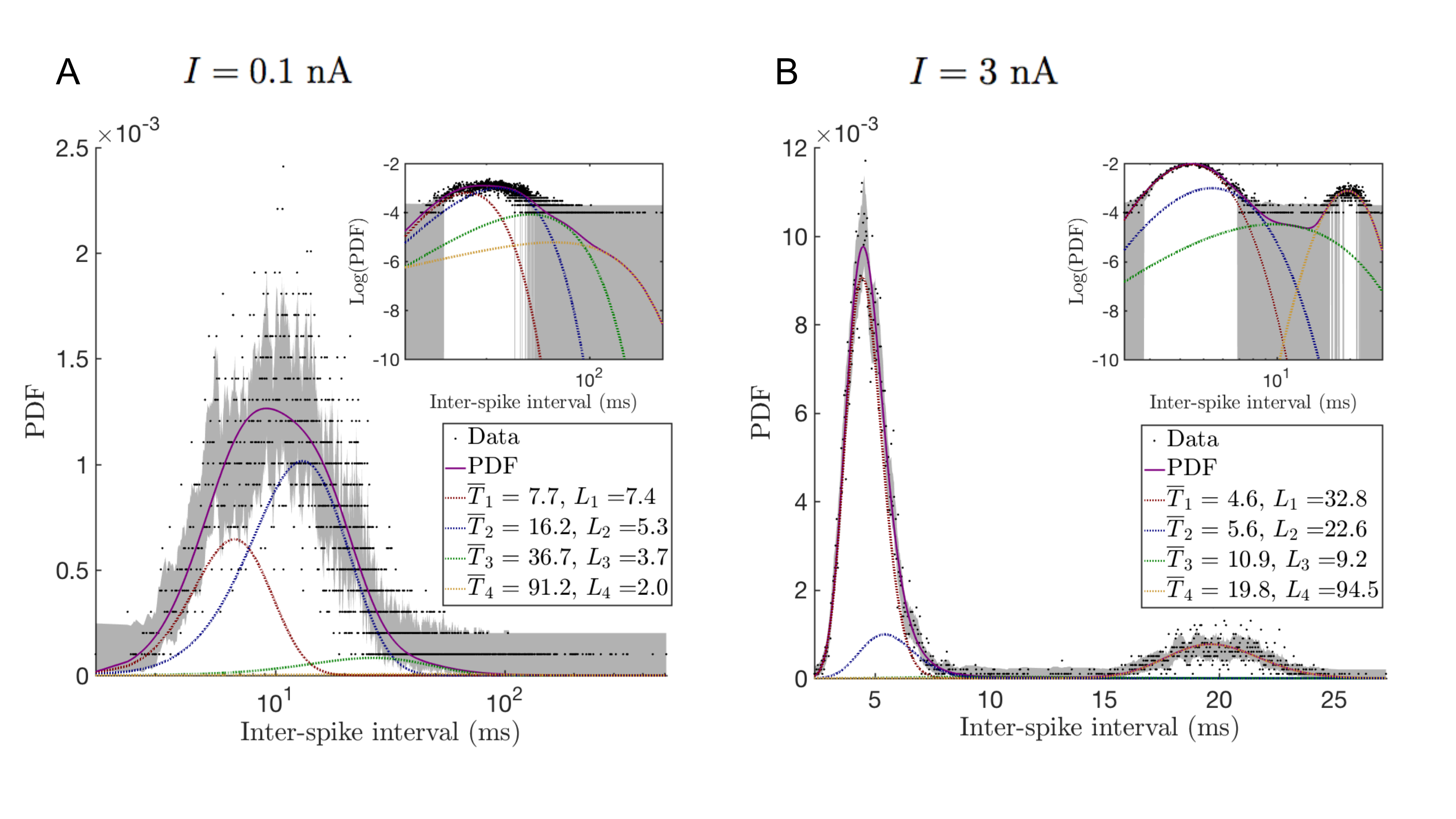}
\caption{{\bf Decomposition of the completion time PDF into contributions from different paths} for (A) $I=0.1$ nA and (B) $I=3.0$ nA. Insets show the same data in log-log units. In (A), the two pathways with the shortest completion time explain the bulk of the distribution while the pathway with the longest average completion time approximate the left tail of the distribution.   In (B), pathways with shortest/longest completion time contribute mostly to the intra/inter burst time scales.}
\label{S3_Fig}
\end{center}
\end{figure}


%
%
%

\bibliography{main}

\begin{thebibliography}{10}

\bibitem{belIlya2009}
Bel G, Munsky B, Nemenman I.
\newblock The simplicity of completion time distributions for common complex
  biochemical processes.
\newblock Physical biology. 2009;7(1):016003.

\bibitem{gutenkunst2007universally}
Gutenkunst RN, Waterfall JJ, Casey FP, Brown KS, Myers CR, Sethna JP.
\newblock Universally sloppy parameter sensitivities in systems biology models.
\newblock PLoS Comput Biol. 2007;3(10):e189.

\bibitem{sinitsyn2009adiabatic}
Sinitsyn N, Hengartner N, Nemenman I.
\newblock Adiabatic coarse-graining and simulations of stochastic biochemical
  networks.
\newblock Proceedings of the National Academy of Sciences.
  2009;106(26):10546--10551.

\bibitem{machta2013parameter}
Machta BB, Chachra R, Transtrum MK, Sethna JP.
\newblock Parameter space compression underlies emergent theories and
  predictive models.
\newblock Science. 2013;342(6158):604--607.

\bibitem{transtrum2015perspective}
Transtrum MK, Machta BB, Brown KS, Daniels BC, Myers CR, Sethna JP.
\newblock Perspective: Sloppiness and emergent theories in physics, biology,
  and beyond.
\newblock The Journal of chemical physics. 2015;143(1):07B201\_1.

\bibitem{borisov2008domain}
Borisov NM, Chistopolsky AS, Faeder JR, Kholodenko BN.
\newblock Domain-oriented reduction of rule-based network models.
\newblock IET systems biology. 2008;2(5):342--351.

\bibitem{hlavacek2006rules}
Hlavacek WS, Faeder JR, Blinov ML, Posner RG, Hucka M, Fontana W.
\newblock Rules for modeling signal-transduction systems.
\newblock Sci STKE. 2006;2006(344):re6--re6.

\bibitem{chylek2015modeling}
Chylek LA, Harris LA, Faeder JR, Hlavacek WS.
\newblock Modeling for (physical) biologists: an introduction to the rule-based
  approach.
\newblock Physical biology. 2015;12(4):045007.

\bibitem{conzelmann2008exact}
Conzelmann H, Fey D, Gilles ED.
\newblock Exact model reduction of combinatorial reaction networks.
\newblock BMC systems biology. 2008;2(1):78.

\bibitem{munsky2006finite}
Munsky B, Khammash M.
\newblock The finite state projection algorithm for the solution of the
  chemical master equation.
\newblock The Journal of chemical physics. 2006;124(4):044104.

\bibitem{haseltine2002approximate}
Haseltine EL, Rawlings JB.
\newblock Approximate simulation of coupled fast and slow reactions for
  stochastic chemical kinetics.
\newblock The Journal of chemical physics. 2002;117(15):6959--6969.

\bibitem{kim2017reduction}
Kim JK, Sontag ED.
\newblock Reduction of multiscale stochastic biochemical reaction networks
  using exact moment derivation.
\newblock PLoS computational biology. 2017;13(6):e1005571.

\bibitem{kang2013separation}
Kang HW, Kurtz TG, et~al.
\newblock Separation of time-scales and model reduction for stochastic reaction
  networks.
\newblock The Annals of Applied Probability. 2013;23(2):529--583.

\bibitem{anderson2011model}
Anderson J, Chang YC, Papachristodoulou A.
\newblock Model decomposition and reduction tools for large-scale networks in
  systems biology.
\newblock Automatica. 2011;47(6):1165--1174.

\bibitem{huang2005systematic}
Huang H, Fairweather M, Griffiths J, Tomlin A, Brad R.
\newblock A systematic lumping approach for the reduction of comprehensive
  kinetic models.
\newblock Proceedings of the Combustion Institute. 2005;30(1):1309--1316.

\bibitem{rao2014model}
Rao S, Van~der Schaft A, Van~Eunen K, Bakker BM, Jayawardhana B.
\newblock A model reduction method for biochemical reaction networks.
\newblock BMC systems biology. 2014;8(1):52.

\bibitem{maurya2005reduced}
Maurya M, Bornheimer S, Venkatasubramanian V, Subramaniam S.
\newblock Reduced-order modelling of biochemical networks: application to the
  GTPase-cycle signalling module.
\newblock IEE Proceedings-Systems Biology. 2005;152(4):229--242.

\bibitem{redner2001guide}
Redner S.
\newblock A guide to first-passage processes.
\newblock Cambridge University Press; 2001.

\bibitem{iyer2016first}
Iyer-Biswas S, Zilman A.
\newblock First-Passage Processes in Cellular Biology.
\newblock Advances in Chemical Physics. 2016;160:261--306.

\bibitem{chou2014first}
Chou T, D'Orsogna MR.
\newblock First passage problems in biology.
\newblock In: First-passage phenomena and their applications. World Scientific;
  2014. p. 306--345.

\bibitem{bressloff2013stochastic}
Bressloff PC, Newby JM.
\newblock Stochastic models of intracellular transport.
\newblock Reviews of Modern Physics. 2013;85(1):135.

\bibitem{zhang2016first}
Zhang Y, Dudko OK.
\newblock First-passage processes in the genome.
\newblock Annual review of biophysics. 2016;45:117--134.

\bibitem{raj2008nature}
Raj A, van Oudenaarden A.
\newblock Nature, nurture, or chance: stochastic gene expression and its
  consequences.
\newblock Cell. 2008;135(2):216--226.

\bibitem{bressloff2014stochastic}
Bressloff PC.
\newblock Stochastic processes in cell biology. vol.~41.
\newblock Springer; 2014.

\bibitem{munsky2012using}
Munsky B, Neuert G, Van~Oudenaarden A.
\newblock Using gene expression noise to understand gene regulation.
\newblock Science. 2012;336(6078):183--187.

\bibitem{wallace1958asymptotic}
Wallace DL.
\newblock Asymptotic approximations to distributions.
\newblock The Annals of Mathematical Statistics. 1958;29(3):635--654.

\bibitem{daniels2015}
Daniels BC, Nemenman I.
\newblock Automated adaptive inference of phenomenological dynamical models.
\newblock Nature communications. 2015;6:8133.

\bibitem{schwarz1978estimating}
Schwarz G, et~al.
\newblock Estimating the dimension of a model.
\newblock The annals of statistics. 1978;6(2):461--464.

\bibitem{kass1995bayes}
Kass RE, Raftery AE.
\newblock Bayes factors.
\newblock Journal of the american statistical association.
  1995;90(430):773--795.

\bibitem{chipman2001practical}
Chipman H, George EI, McCulloch RE, Clyde M, Foster DP, Stine RA.
\newblock The practical implementation of Bayesian model selection.
\newblock Lecture Notes-Monograph Series. 2001;p. 65--134.

\bibitem{rissanen1999hypothesis}
Rissanen J.
\newblock Hypothesis selection and testing by the MDL principle.
\newblock The Computer Journal. 1999;42(4):260--269.

\bibitem{balasubramanian1997statistical}
Balasubramanian V.
\newblock Statistical inference, Occam's razor, and statistical mechanics on
  the space of probability distributions.
\newblock Neural computation. 1997;9(2):349--368.

\bibitem{mackay2003information}
MacKay DJ, Mac~Kay DJ.
\newblock Information theory, inference and learning algorithms.
\newblock Cambridge university press; 2003.

\bibitem{Kolomeisky}
Li X, Kolomeisky AB.
\newblock Mechanisms and topology determination of complex chemical and
  biological network systems from first-passage theoretical approach.
\newblock The Journal of chemical physics. 2013;139(14):10B606\_1.

\bibitem{tuckwell1988}
Tuckwell HC.
\newblock Introduction to theoretical neurobiology: volume 2, nonlinear and
  stochastic theories.
\newblock Cambridge University Press; 1988.

\bibitem{Correia1977}
Correia M, Landolt J.
\newblock A point process analysis of the spontaneous activity of anterior
  semicircular canal units in the anesthetized pigeon.
\newblock Biological cybernetics. 1977;27(4):199--213.

\bibitem{Rodieck1962}
Rodieck R, Kiang NS, Gerstein G.
\newblock Some quantitative methods for the study of spontaneous activity of
  single neurons.
\newblock Biophysical Journal. 1962;2(4):351--368.

\bibitem{Grossman1961}
Grossman R, Viernstein L.
\newblock Discharge patterns of neurons in cochlear nucleus.
\newblock Science. 1961;134(3472):99--101.

\bibitem{Lamarre1971}
Lamarre Y, Filion M, Cordeau J.
\newblock Neuronal discharges of the ventrolateral nucleus of the thalamus
  during sleep and wakefulness in the cat I. Spontaneous activity.
\newblock Experimental brain research. 1971;12(5):480--498.

\bibitem{Steriade1973}
Steriade M, Wyzinski P, Apostol V.
\newblock Differential synaptic reactivity of simple and complex pyramidal
  tract neurons at various levels of vigilance.
\newblock Experimental brain research. 1973;17(1):87--110.

\bibitem{Tolhurst1981}
Tolhurst D, Movshon JA, Thompson I.
\newblock The dependence of response amplitude and variance of cat visual
  cortical neurones on stimulus contrast.
\newblock Experimental brain research. 1981;41(3-4):414--419.

\bibitem{Bishop1964}
Bishop P, Levick W, Williams W.
\newblock Statistical analysis of the dark discharge of lateral geniculate
  neurones.
\newblock The Journal of physiology. 1964;170(3):598--612.

\bibitem{Burns1976}
Burns BD, Webb A.
\newblock The spontaneous activity of neurones in the cat’s cerebral cortex.
\newblock Proc R Soc Lond B. 1976;194(1115):211--223.

\bibitem{Tuckwell1978}
Tuckwell HC, Richter W.
\newblock Neuronal interspike time distributions and the estimation of
  neurophysiological and neuroanatomical parameters.
\newblock Journal of theoretical biology. 1978;71(2):167--183.

\bibitem{bair1994power}
Bair W, Koch C, Newsome W, Britten K.
\newblock Power spectrum analysis of bursting cells in area MT in the behaving
  monkey.
\newblock Journal of Neuroscience. 1994;14(5):2870--2892.

\bibitem{debusk1997stimulus}
DeBusk B, DeBruyn E, Snider R, Kabara J, Bonds A.
\newblock Stimulus-dependent modulation of spike burst length in cat striate
  cortical cells.
\newblock Journal of Neurophysiology. 1997;78(1):199--213.

\bibitem{nowak2003electrophysiological}
Nowak LG, Azouz R, Sanchez-Vives MV, Gray CM, McCormick DA.
\newblock Electrophysiological classes of cat primary visual cortical neurons
  in vivo as revealed by quantitative analyses.
\newblock Journal of neurophysiology. 2003;89(3):1541--1566.

\bibitem{shih2011improved}
Shih JY, Atencio CA, Schreiner CE.
\newblock Improved stimulus representation by short interspike intervals in
  primary auditory cortex.
\newblock Journal of neurophysiology. 2011;105(4):1908--1917.

\bibitem{tsubo2012power}
Tsubo Y, Isomura Y, Fukai T.
\newblock Power-law inter-spike interval distributions infer a conditional
  maximization of entropy in cortical neurons.
\newblock PLoS Comput Biol. 2012;8(4):e1002461.

\bibitem{sungho2016}
Hong S, Negrello M, Junker M, Smilgin A, Thier P, De~Schutter E.
\newblock Multiplexed coding by cerebellar Purkinje neurons.
\newblock Elife. 2016;5.

\bibitem{Nemenman2005}
Nemenman I.
\newblock {Fluctuation-dissipation theorem and models of learning.}
\newblock Neural Comput. 2005;17.

\bibitem{tierney1986accurate}
Tierney L, Kadane JB.
\newblock Accurate approximations for posterior moments and marginal densities.
\newblock Journal of the american statistical association. 1986;81(393):82--86.

\bibitem{geweke1989bayesian}
Geweke J.
\newblock Bayesian inference in econometric models using Monte Carlo
  integration.
\newblock Econometrica: Journal of the Econometric Society. 1989;p. 1317--1339.

\bibitem{smith1993bayesian}
Smith AF, Roberts GO.
\newblock Bayesian computation via the Gibbs sampler and related Markov chain
  Monte Carlo methods.
\newblock Journal of the Royal Statistical Society: Series B (Methodological).
  1993;55(1):3--23.

\bibitem{ding2018model}
Ding J, Tarokh V, Yang Y.
\newblock Model selection techniques: An overview.
\newblock IEEE Signal Processing Magazine. 2018;35(6):16--34.

\bibitem{Owen2013}
Owen AB.
\newblock Monte Carlo theory, methods and examples; 2013.

\bibitem{DeSchutter1994}
De~Schutter E, Bower JM.
\newblock {An active membrane model of the cerebellar Purkinje cell I.
  Simulation of current clamps in slice}.
\newblock Neurophysiology. 1994;71.

\bibitem{Miyasho2000}
Miyasho T, Takagi H, Suzuki H, Watanabe S, Inoue M, Kudo Y, et~al.
\newblock Low-threshold potassium channels and a low-threshold calcium channel
  regulate Ca2+ spike firing in the dendrites of cerebellar Purkinje neurons: a
  modeling study.
\newblock Brain research. 2001;891(1-2):106--115.

\bibitem{santamaria2002modulatory}
Santamaria F, Jaeger D, De~Schutter E, Bower JM.
\newblock Modulatory effects of parallel fiber and molecular layer interneuron
  synaptic activity on Purkinje cell responses to ascending segment input: a
  modeling study.
\newblock Journal of computational neuroscience. 2002;13(3):217--235.

\bibitem{kulagina2007electro}
Kulagina IB, Korogod SM, Horcholle-Bossavit G, Batini C, Tyc-Dumont S.
\newblock The electro-dynamics of the dendritic space in Purkinje cells of the
  cerebellum.
\newblock Archives italiennes de biologie. 2007;145(3):211--233.

\bibitem{forrest2012sodium}
Forrest MD, Wall MJ, Press DA, Feng J.
\newblock The sodium-potassium pump controls the intrinsic firing of the
  cerebellar Purkinje neuron.
\newblock PloS one. 2012;7(12):e51169.

\bibitem{nemenman2002entropy}
Nemenman I, Shafee F, Bialek W.
\newblock Entropy and inference, revisited.
\newblock In: Advances in neural information processing systems; 2002. p.
  471--478.

\bibitem{lin1991}
Lin J.
\newblock Divergence Measures Based on the {{Shannon}} Entropy.
  1991;37(1):145--151.

\bibitem{valleriani2014unveiling}
Valleriani A, Li X, Kolomeisky AB.
\newblock Unveiling the hidden structure of complex stochastic biochemical
  networks.
\newblock The Journal of chemical physics. 2014;140(6):02B608\_1.

\end{thebibliography}



\end{document}